\documentclass{iopjournal}
\usepackage[english]{babel}
\usepackage{graphicx}
\usepackage{hyperref}
\usepackage{color}
\usepackage{bm}
\usepackage{soul}
\usepackage{subcaption}
\usepackage{amsmath,amssymb}
\usepackage[final]{microtype}

\usepackage{bm}

\definecolor{arsenic}{rgb}{0.23, 0.27, 0.29}
\definecolor{darkgreen}{rgb}{0.0, 0.2, 0.03}
\definecolor{migris}{rgb}{0.37, 0.37, 0.37}

\usepackage{booktabs}
\usepackage{float} 

\begin{document}
\articletype{paper}

\title{From Dark Radiation to Dark Energy: Unified Cosmological Evolution in K-essence Models}

\author{Eladio Moreno, $^1$\orcid{0000-0002-5400-2584} and Josue De-Santiago $^{2,3,*}$\orcid{0000-0002-1163-3730}}

\affil{$^1$Departamento de F\'isica, Universidad de Guanajuato - DCI, 37150, Le\'on, Guanajuato, M\'exico.}

\affil{$^2$Departamento de Física, Centro de Investigación y de Estudios Avanzados del I.P.N.,
Apartado Postal 14-740, 07000 Ciudad de México, México.}

\affil{$^3$Secretaría de Ciencia, Humanidades, Tecnología e Innovación,  Av.   Insurgentes  Sur  1582,  Colonia  Crédito Constructor, Del.  Benito Juárez, 03940, Ciudad de México, México.}

\affil{$^*$Author to whom any correspondence should be addressed.}

\email{ea.morenoalcala@ugto.mx, Josue.desantiago@cinvestav.mx}

\begin{abstract}

We study a class of Unified Dark Matter (UDM) models based on generalized K-essence, where a single scalar field with non-canonical kinetic terms accounts for dark radiation, dark matter, and dark energy. Starting from the purely kinetic Lagrangian proposed by Scherrer in \cite{Scherrer:2004au}, we extend the analysis to quadratic and exponential scalar potentials and explore their phenomenology. All models are implemented in a modified version of \texttt{Hi\_CLASS} 
(a Horndeski capable extension of CLASS)
and confronted with data from \textit{Planck} 2018, DESI DR1, and Big Bang Nucleosynthesis. The scenarios reproduce the full sequence of cosmic epochs: an early radiation-like phase, a matter-dominated era, and late-time accelerated expansion. The new models predict slightly higher values of the Hubble constant compared to $\Lambda$CDM, thereby partially alleviating the respective tensions from $\sim 4.4 \sigma$ to $\sim 3.4 \sigma$. The quadratic potential requires an ultralight mass that makes it effectively indistinguishable from the Scherrer solution.  Overall, generalized K-essence provides a minimal (single field and minimally coupled) and observationally viable realization of UDM, offering a unified description of the dark sector with distinctive signatures in both early- and late-time cosmology.

\end{abstract}

\section{Introduction}
\label{sec:intro}

The $\Lambda$CDM framework has long served as the reference model in cosmology, providing an excellent fit to many key observations, including the cosmic microwave background (CMB), baryon acoustic oscillations (BAO), and large-scale structure. Its minimal six-parameter set and consistency with early-Universe probes have established it as the prevailing benchmark for precision cosmology. However, recent analyses, particularly those incorporating the latest DESI measurements \cite{DESI:2024mwx}, indicate that the cosmological constant may not fully account for the data. 
Another notable discrepancy in the cosmology measurements is the $H_0$ tension. Which refers to the $\sim 5\sigma$ difference between the Hubble constant inferred from Planck measurements of the CMB~\cite{Planck:2018vyg} and local distance-ladder determinations such as those reported by the SH0ES collaboration~\cite{Riess:2021jrx}. This would also be relaxed by a change in the dark sector behavior \cite{di2024hubble}. This has renewed interest in exploring alternative scenarios capable of addressing emerging tensions between early- and late-time observable                                                                                                                                                                                                                                                                                                                                                                                                                                                                                                                                                                                                                                                                                                                                                                                                                                                                                                                                                                                                                                                                                                                                                                                                                                                                                                                                                                                                                                                                                                                                                                                                                                                                                                                                                                                                                                                                                                                                                                                                                                                                                                                                                                                                                                                                                                                                                                                                                                                                                                                                                                                                                                                                                                                                                                                                                                                                                                                                                                                                                                                                                                                                                                                                                                                                                                                                                                                                                                                                                                                                                                                                                                                                                                                                                                                                      s.

Scalar field theories provide a natural framework for such extensions. Canonical quintessence scenarios~\cite{Ratra:1987rm,Caldwell:1997ii, Ferreira:1997hj} describe dark energy through a scalar potential, while generalized models with non-canonical kinetic terms, known as K-essence~\cite{Armendariz-Picon:1999hyi,Chiba:1999ka,Armendariz-Picon:2000ulo, Armendariz-Picon:2000nqq}, broaden the phenomenology and allow cosmic acceleration to arise from the kinetic structure itself. A minimal and analytically tractable realization is the purely kinetic model proposed by Scherrer~\cite{Scherrer:2004au}, in which the scalar Lagrangian is expanded quadratically around a background value $X_0$. This construction allows the field to behave as radiation at early times, as matter at intermediate epochs, and as vacuum energy at late times, thereby realizing the unified dark matter (UDM) picture. Related approaches have shown that tachyon and Chaplygin-type scenarios~\cite{Chimento:2003ta}, models unifying inflation with dark matter and dark energy~\cite{Bose:2008ew}, and more general K-essence functions with a minimum~\cite{De-Santiago:2011aka} all fall within this broad class, underscoring its generality and robustness~\cite{Hussain:2024qrd}.

Despite its interesting properties, a purely kinetic model reduces to a barotropic fluid model which reduces the rich phenomenology of a field theory of the dark sector \cite{PhysRevD.72.063502}. In this work, we revisit the Scherrer's model and consider two extensions that incorporate scalar potentials: a quadratic potential and an exponential one, that are introduced as additive terms in the Lagrangian as adopted previous works~\cite{Bertacca:2007ux, Bertacca:2010ct, De-Santiago:2012ibi}. These scenarios allow us to test if potentials can provide a richer phenomenology  or leave distinctive observational imprints.

Our analysis has three main objectives: (i) to test the observational viability of the Scherrer model and its potential-extended generalizations; (ii) to quantify their impact on key cosmological parameters, including
the physical dark matter density $\omega_{\rm dm}$, the present Hubble expansion rate $H_0$, and the clustering parameter $S_8\equiv \sigma_8 \left( \Omega_m/{0.3} \right)^{1/2}$;
and (iii) to identify distinctive features in the background and perturbation dynamics, with particular emphasis on the early-time relativistic phase of the scalar field and its imprint on 
the  effective number of relativistic species
$N_{\rm eff}$. These results are then compared with current data and critically assessed in light of their ability to address the $H_0$ and $S_8$ tensions. 

The remainder of this paper is organized as follows. In Section~\ref{sec:k-essence} we introduce the theoretical framework of generalized K-essence as a realization of Unified Dark Matter, starting from the Sherrer solution and extending it with quadratic and exponential potentials. Section~\ref{sec:data} describes the observational datasets and the numerical  methodology, including the implementation of the models in \texttt{Hi\_CLASS} and the parameter inference with \texttt{MontePython}. Our main results are presented in Section~\ref{sec:results_discussion}, where we discuss the background evolution, perturbation behavior, and cosmological constrainst for the different scenarios. Finally, Section~\ref{sec:conclusions} summarizes our findings, assesses their implications for current cosmological tensions, and outlines possible directions for future work.

\section{K-Essence Models in Cosmology}
\label{sec:k-essence}

K-essence theories provide a general class of scalar field models characterized by non-canonical kinetic terms. Originally developed to drive inflation~\cite{Armendariz-Picon:1999hyi}, they were later extended to describe dark energy and unified dark sector dynamics~\cite{Chiba:1999ka, Armendariz-Picon:2000ulo}. %

The action for a minimally coupled K-essence field reads:
\begin{equation}
\mathcal{L} = \int d^4x \sqrt{-g} \left( \frac{R}{2\kappa^2} + G_2(X,\phi) + \mathcal{L}_m \right),
\end{equation}
where $G_2(X,\phi)$ is the scalar field Lagrangian, $\mathcal{L}_m$ corresponds to standard matter, and $\kappa^2 = 8\pi G$. We assume a functional form
\begin{equation}\label{full_lagrangian}
    G_2(X,\phi) = F(X) - V(\phi),
\end{equation}
with \(X \equiv -\tfrac{1}{2} g^{\mu\nu} \partial_\mu \phi \partial_\nu \phi\).  
This decomposition is widely adopted in scalar field cosmologies and effective field theory approaches~\cite{Bertacca:2007ux, Bertacca:2010ct, De-Santiago:2012ibi,PhysRevD.67.123503,PhysRevD.68.043509}.

This field has an energy-momentum tensor  of:
\begin{equation}
T^\mu_{\ \nu} = G_{2,X} \partial^\mu \phi \partial_\nu \phi + \delta^\mu_{\ \nu} G_2,
\end{equation}
with energy density and pressure:
\begin{equation}\label{rho_ano_p}
\rho_\phi = -G_2 + 2X G_{2,X}, \qquad P_\phi = G_2.
\end{equation}

In a flat Friedmann-Lamaitre-Robertson-Walker (FLRW) background, the field evolution is governed by the generalized Klein-Gordon equation:
\begin{equation}\label{equation::k-g}
\left( F_X + 2X F_{XX} \right) \phi'' + 2\mathcal{H} \left( F_X - X F_{XX} \right) \phi' = -a^2 V_\phi,
\end{equation}
where primes denote conformal time derivatives and $\mathcal{H} = a'/a$.

\subsection{Purely Kinetic K-Essence}

In the absence of a scalar potential, the dynamics of the field are governed entirely by the non-canonical kinetic term $F(X)$. 
For purely kinetic models, the Lagrangian is invariant under constant shifts of the scalar field, $\phi \to \phi + \phi_0$, implying the existence of a conserved Noether current associated with this shift symmetry satisfying the equation $(\sqrt{-g}F_X\partial^\mu \phi)_{,\mu}=0$. In a background FLRW this leads to
\begin{equation}\label{noether}
X F_X^2 = k a^{-6},
\end{equation}
where $k$ is a constant of integration.
This result assumes homogeneity and isotropy of the background, the absence of an explicit potential term, and no explicit dependence of the Lagrangian on the spacetime coordinates.
This describes how $X$ evolves as the Universe expands. Relaxing any of these assumptions, for instance by introducing a scalar potential or explicit time dependence, generically breaks the conservation law and modifies the background evolution. In section \ref{sec:potentials} and in subsequent sections we will explore solutions to the field in the presence of a scalar potential with numerical methods as equation \eqref{noether} will no longer be valid.

If $F(X)$ has a minimum at a certain value $X_0$, it can be approximated as a quadratic function close to this minimum. This led Scherrer in \cite{Scherrer:2004au} to introduce the purely kinetic quadratic model
\begin{equation}\label{scherrer}
F(X) = -F_0 + F_2 (X - X_0)^2,
\end{equation}
where $F_0$, $F_2$, and $X_0$ are constants. 
In the regime where $(X - X_0)/X_0 \ll 1$,
the kinetic term $X$ evolves as:
\begin{equation}\label{approx}
 X = X_0 \left[ 1 + \frac{\theta}{a^{3}} \right]\,,
\end{equation}
where $\theta$ is a constant dependent on the initial conditions of the
field, and $\theta/a^3\ll1$ for this approximation to be valid. The corresponding energy density takes the form
\begin{equation}\label{scherrer_density}
\rho_\phi \simeq F_0 + 4 F_2 X_0^2  \frac{\theta}{a^3} ,
\end{equation}
which represents a superposition of a cosmological constant and a pressureless matter component.  
Identifying these contributions with present-day density parameters, we obtain
\begin{equation}
F_0 = \Omega_{F_0} \, \rho_c^{(0)}, 
\qquad 
F_2 = \frac{\Omega_{\rm dm} \, \rho_c^{(0)}}{4 X_0^2 \theta}, \label{relationF2_F0}
\end{equation}
where $\Omega_{F_0}$ denotes the present-day density fraction associated with the constant term (effectively playing the role of dark energy), and $\rho_c^{(0)} = 3H_0^2/(8\pi G)$ is the critical density today.
Once the present day dark sector density is fixed, the late-time background evolution is primarily controlled by the physically independent combinations $(\Omega_{\rm dm},\Omega_{F_0})$, while $\theta$ sets the duration of the early-time radiation-like phase and determines when the solution transitions into the regime $(X-X_0)/X_0\ll 1$.

\subsection{Early-Time Evolution and BBN Constraints}\label{sec:early}

In the early Universe, when the scale factor satisfies \( a \ll 1 \), the kinetic term is much larger than its present value, i.e., \( X \gg X_0 \).
In order to obtain constrains to the field from this epoch we will assume from now on that the
quadratic Lagrangian \eqref{scherrer} is exact for all the relevant scales and not
only an approximation for $X$ close to $X_0$
(See \cite{Csillag:2025gnz} for a theoretical motivation of the quadratic kinetic Lagrangian). As equation \eqref{noether} is exact, substituting $F_X$ we obtain
\begin{equation}\label{third}
    (X-X_0)X=\frac{\theta^2X_0^3}{a^{6}}\,,
\end{equation}
where we have used the relation $k=4\theta^2X_0^3F_2^2$ to substitute $k$ in terms of the constant $\theta$ used in \eqref{approx}.
\begin{figure}[H]
\centering
\includegraphics[width=0.6\linewidth]{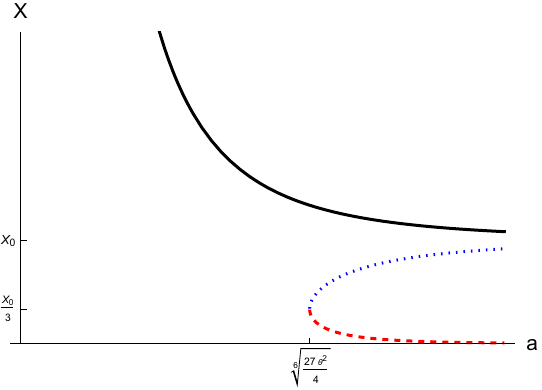}
\caption{Solutions to $X(a)$ to the purely kinetic Lagrangian. As the evolution equation \eqref{third} is cubic, it has three different solutions. Two of which are only valid for $a>\sqrt[6]{27\theta^2/4}$ (red dashed and blue dotted curves), and one valid for all $a>0$ (black continuous curve). Assuming that the field starts evolving with $X\ll X_0$ at $a\ll 1$ the physical solution is the black continuous line given by the expression \eqref{sol_cual}.}
\label{fig:exact_sol}
\end{figure}
This expression is a cubic equation in $X$ with 3 solutions, but we are interested in the one for $X>X_0$ which, in turn, is the only one valid for all $a$ given by
\begin{equation}
    X=\frac{1}{6} {X_0} \left(\frac{\left(6 \sqrt{81 \theta ^4-12 a^6 \theta ^2}-4 a^6+54 \theta ^2\right)^{2/3}+2 \sqrt[3]{2} a^4}{a^2 \sqrt[3]{3 \sqrt{81 \theta ^4-12 a^6 \theta ^2}-2 a^6+27 \theta ^2}}+4\right)\,.
    \label{sol_cual}
\end{equation}

We remark that the solution \eqref{sol_cual} is valid for all the values of $X>X_0$ if the Lagrangian is purely kinetic and exactly quadratic. 
Under the approximation of $a\ll 1$, the asymptotic behavior of the solution becomes:
\begin{equation}\label{early_aproximation}
X(a) \approx  \frac{X_0\theta^{2/3}}{a^2}.
\end{equation}

In this limit, the scalar field dynamics are dominated by the quadratic kinetic term. The energy density then scales as $\rho_\phi \propto a^{-4}$, mimicking the behavior of a radiation fluid.

During the early Universe, the scalar field behaves as a radiation-like component with equation of state $w_\phi = 1/3$. In this regime, it contributes to the total radiation density and thus to the parameter $N_{\mathrm{eff}}$. The latter is tightly constrained by primordial nucleosynthesis and the CMB, which are sensitive to any additional relativistic degrees of freedom photons and the three neutrino families. Therefore, it is essential to quantify the scalar contribution to $N_{\mathrm{eff}}$ in order to ensure that the radiation-like density remains compatible with BBN and recombination constraints.

Defining the total relativistic density of the universe as proportional to the
photon density $\rho_\gamma$, gives
\begin{equation}
\rho_{\text{rel}} = \left[1 + \frac{7}{8}\left(\frac{4}{11}\right)^{4/3} N_{\text{eff}} \right] \rho_\gamma,
\end{equation}
with $N_{\mathrm{eff}}\simeq 3.046$ for the 3 standard model neutrino flavors in the
absence of other relativistic components
\cite{de2016relic, abenza2020precision, akita2020precision, froustey2020neutrino}. In our case, the scalar field
adds an extra component to the $N_{\rm{eff}}$ as 
\begin{equation}\label{neff_completa}
\Delta N_{\mathrm{eff}} = \frac{8}{7} \left( \frac{11}{4} \right)^{4/3} 
\left( \frac{3P_\phi}{\rho_\gamma} \right),
\end{equation}
This expression makes explicit how the scalar field modifies $N_{\mathrm{eff}}$ during the radiation era. 
From the solution (\ref{early_aproximation}) and using $F_2$ from eq.
(\ref{relationF2_F0}), we see that
\begin{equation}\label{bound_neff}
    \Delta N_{\mathrm{eff}} \simeq \frac{6}{7}\left( \frac{11}{4}\right)^{4/3}
    \frac{\Omega_{\rm{dm}}\theta^{1/3}}{\Omega_\gamma} \,,
\end{equation}
which explicitly depends on the initial conditions parameter $\theta$.
In section \ref{sec:background}, we will use
this modification on the relativistic degrees of freedom  to constrain the model
against observations, where we will use a more precise numerical solution
and the expression (\ref{neff_completa}) to determine $N_{\mathrm{eff}}$.

\subsection{Extensions with Scalar Potentials}\label{sec:potentials}
Different cosmological works have treated the separable Lagrangian form \eqref{full_lagrangian} with a non-zero potential. In particular, 
Refs. \cite{Bose:2008ew,De-Santiago:2011aka,fang2007cosmologies,mukhanov2006enhancing,vikman2006inflation}
investigated quadratic potentials in the context of non-canonical fields where the potential drives inflation in the early universe.%
\footnote{In the context of canonical scalar fields quadratic potentials arise naturally from effective field theory. They have been widely employed to model inflation \cite{linde1983chaotic}, dark energy \cite{PhysRevLett.95.141301}, and dark matter \cite{PhysRevD.63.063506}.}
Our focus here is in the post inflationary Universe, and we are interested to see the impact of this term on the subsequent evolution of the field. We therefore introduce the quadratic potential
\begin{equation}
    V_{m}(\phi) = \frac{1}{2} m_\phi^2 \phi^2\,. \label{quadratic}
\end{equation}
However, as we will show in section \ref{sec:constraints}, the non-canonical kinetic sector drives the field towards large values at late times, which can significantly alter the cosmological dynamics if the potential grows with $\phi$. This motivates us to consider alternative potential that vanish at large field values. In this spirit, following \cite{copeland1998exponential,De-Santiago:2012ibi}, we propose an exponentially decreasing potential of the form
\begin{equation}
V_{\rm{exp}}(\phi) = V_0 e^{-\lambda \phi} \,. \label{exponential}
\end{equation}

From the full Klein-Gordon equation, the evolution of $X$ obeys:
\begin{equation}\label{background}
X' = -\frac{\phi' V_\phi + 6 \mathcal{H} X F_X}{F_X + 2X F_{XX}} \,,
\end{equation}
which now its not solvable as the purely kinetic model. In order obtain
cosmological predictions from this model and to constrain
its parameters, in the next section we will solve equation (\ref{background})
numerically with the help of \texttt{Hi\_CLASS}~\cite{Blas:2011rf,Zumalacarregui:2016pph, Bellini:2019syt}. It will solve the background field equation
(\ref{background}) with the condition that $\rho_\phi$ today be equal to the
dark sector component
\begin{equation}\label{target}
    \frac{\rho_\phi^{(0)}}{\rho_{\rm{crit}}^{(0)}} = \Omega_\Lambda + \Omega_{dm} \,.
\end{equation}
To satisfy this condition the code performs a shooting method where the first
step is approximated by the purely kinetic early universe conditions
\begin{equation}
X' = \frac{ -\phi' V_\phi / F_2 - 12 \mathcal{H} X(X - X_0) }{6X - 2X_0}\,.
\label{shooting}
\end{equation}
where $F_2$ come from equation (\ref{relationF2_F0}). 
This expression comes from the general Klein--Gordon equation (\ref{background}). And the subsequent iterations modify the initial conditions of the field until the condition (\ref{target}) is satisfied.

\subsection{Phase space analysis and stability}

Once we fix the parameters of the Lagrangian, the evolution of the purely kinetic field has only one degree of freedom which is the integration constant $\theta$, with only one solution valid throughout the entire life of the Universe given by Eq. \eqref{sol_cual} and shown in Fig. \ref{fig:exact_sol}. This solution ensures the asymptotic behaviors of early Universe radiation and late Universe matter + cosmological constant studied in the previous sections. In this section we will show that those asymptotic behaviors still arise in the presence of scalar potentials under certain generic initial conditions. To show this we will introduce two dynamical systems valid one at the early Universe and the other at the late Universe.

Moreover, in this section we will explore the possibility that the field would enter a region of phase space with a ghost instability $G_{2,X}=2F_2(X-X_0)<0$ or gradient instability $c_s^2=(X-X_0)/(3X-X_0)<0$. Both can only occur if the field reaches the region with $X<X_0$. The solution \eqref{sol_cual}  already protects the purely kinetic model from entering this region, and in this section we will show that the exponential model is also  protected if it starts with $\dot \phi>0$, while the  quadratic model can enter this region and become unstable in the future evolution of the field.

For the early Universe  analysis we  will base our results on Ref. \cite{De-Santiago:2012ibi}, where the authors studied the phase space of a
Universe filled with a scalar field---with a Lagrangian of the type \eqref{full_lagrangian}---and a barotropic fluid. Here we will assume that the barotropic fluid has a equation of state $w_r=1/3$ corresponding to the radiation (photons, neutrinos) in the early Universe. Defining the dynamical variables
\begin{equation}
    x =\sqrt{\frac{2XF_X-F}{\rho_c}}\,,\qquad y=\sqrt{\frac{V}{\rho_c}}\,,
\end{equation}
where we will consider that $X$ dominates over the constants at the early Universe, approximating the kinetic term as
\begin{equation}\label{approx_lagrangian}
    F(X) = F_2 X^2\,.
\end{equation}
The dynamical system for the newly defined variables reduces to:
\begin{eqnarray}
    \frac{dx}{dN} &=&y^2\left( -2x + \frac{3}{2}\sigma\right)\,,
    \nonumber\\ \label{dydn}
    \frac{dy}{dN} &=&y\left( 2(1-y^2) - 
    \frac{3}{2}\sigma x\right)\,,
\end{eqnarray}
where $dN=d\log(a)$ and
\begin{eqnarray}
    \sigma = -\frac{4M_{\rm{Pl}}}{3\sqrt{F_2}} \frac{1}{\dot \phi \phi} &\qquad& \text{for the quadratic potential,}
    \label{sigma_cuad}\\ \label{sigma_exp}
    \sigma = -\frac{2M_{\rm{Pl}}}{3\sqrt{F_2}} \frac{\lambda}{\dot \phi }
    &\qquad& \text{for the exponential potential.}
\end{eqnarray}

The dynamical system   \eqref{dydn} has equilibrium points along the line $y=0$ with $x\in[0,1]$.
\footnote{The point with $x=3\sigma/4$, $y^2=1-3\sigma/4$ also
cancels the evolution equations but only instantaneously as $\sigma$ evolves in time; see Ref. \cite{De-Santiago:2012ibi}}
Along this line, the field density is purely kinetic and with an equation of state $w_\phi=1/3$ as the one studied in section \ref{sec:early}. The densities of the field and the radiation evolve at the same rate, with $x^2$ the fraction of the total density of the Universe stored in the field. This behavior resembles that of scaling solutions, with the restriction that the critical points here do not admit other equations of state different from $1/3$.

The  stability of this line follows from the behavior of small deviations around $y=0$. When $3x\sigma <4$ the line acts as a past attractor. In this case, the early Universe scenario studied in this work represents a typical state from which the phase space trajectories originate. This restriction is satisfied, for example, when $\sigma<0$, which happens if $\phi$ and $\dot \phi$ share the same sign in the quadratic potential model, or if $\dot \phi>0$ in the exponential potential. Therefore, the early time scenario studied in Sec.~\ref{sec:early} and the numerical background evolution presented in Sec.~\ref{sec:background} is not an isolated solution but a general behavior of the dynamical system.

We now introduce a complementary autonomous system tailored to the late Universe, where we will use the full form of the non canonical kinetic term \eqref{scherrer} and assume that the scalar field dominates the energy content of the Universe.
Our construction follows the same dynamical systems philosophy adopted for non-canonical scalar fields in Ref.~\cite{josue}. Defining the dimensionless variables
\begin{equation}
v \equiv \frac{\dot\phi}{\sqrt{2X_0}},\qquad
\varphi \equiv \frac{H_*}{\sqrt{2X_0}}\,\phi,\qquad
h(\varphi,v)\equiv \frac{H}{H_*},
\label{eq:varphi_v_def_clean}
\end{equation}
where $H_*$ is a fixed reference scale chosen bellow.
We also introduce the positive constant
\begin{equation}
K \equiv F_2 X_0^2-F_0,
\label{eq:K_positive}
\end{equation}
so that the scalar dominated Friedmann equation implies the natural choice
\begin{equation}
H_*^2 \equiv \frac{\kappa^2}{3}K\,.
\label{eq:Hstar_choice}
\end{equation}
The autonomous system follows from the homogeneous Klein--Gordon equation \eqref{equation::k-g} together with the Friedmann equation, written entirely in the variables \eqref{eq:varphi_v_def_clean}.
In the absence of extra matter components the dynamics close as
\begin{equation}
\frac{d\varphi}{dN}=\frac{v}{h(\varphi,v)},\qquad
\frac{dv}{dN}=-3v\,\frac{v^2-1}{3v^2-1}+\mathcal{S}(\varphi,v),
\label{eq:system_general_clean}
\end{equation}
with $h^2$ and $\mathcal{S}$ determined by the choice of potential. For all three choices of potential we introduce the common dimensionless coupling
\begin{equation}
\alpha \equiv \frac{F_2X_0^2}{K},
\label{eq:alpha_def}
\end{equation}
and additional dimensionless parameters that depend on $V(\phi)$:
\begin{itemize}
\item Purely kinetic sector $V(\phi)=0$: 
\begin{equation}
h^2(\varphi,v)=-1+\alpha\,(3v^4-2v^2),\qquad \mathcal{S}=0.
\label{eq:h2_kin_clean}
\end{equation}
\item Quadratic potential $V(\phi)=\tfrac12 m_\phi^2\phi^2$: defining
\begin{equation}
\beta \equiv \frac{3m_\phi^2X_0}{\kappa^2K^2},\qquad
\delta \equiv \frac{3m_\phi^2}{2\kappa^2F_2X_0K},
\label{eq:alpha_beta_delta_def}
\end{equation}
the system becomes
\begin{equation}
h^2(\varphi,v)=-1+\alpha\,(3v^4-2v^2)+\beta\,\varphi^2,\qquad
\mathcal{S}(\varphi,v)=-\,\delta\,\frac{\varphi}{h(\varphi,v)\,(3v^2-1)}.
\label{eq:h2_quad_clean}
\end{equation}
\item Exponential potential $V(\phi)=V_0 e^{-\lambda \phi}$: defining
\begin{equation}
\xi \equiv \frac{V_0}{K},\qquad
\gamma \equiv \lambda\,\frac{\sqrt{2X_0}}{H_*},\qquad
\varepsilon \equiv \frac{\gamma\,\xi}{4\alpha},
\label{eq:alpha_xi_gamma_eps_def}
\end{equation}
one obtains
\begin{equation}
h^2(\varphi,v)=-1+\alpha\,(3v^4-2v^2)+\xi\,e^{-\gamma\varphi},\qquad
\mathcal{S}(\varphi,v)=\varepsilon\,\frac{e^{-\gamma\varphi}}{h(\varphi,v)\,(3v^2-1)}.
\label{eq:h2_exp_clean}
\end{equation}
\end{itemize}

For the exponential potential an alternative definition of the dynamical system arrives when we compactify the direction $\varphi\to +\infty$ by introducing
\begin{equation}
u\equiv e^{-\gamma\varphi}\ge 0,
\label{eq:u_def}
\end{equation}
so that $u\to 0$ corresponds to the asymptotic regime where the exponential potential switches off. In the variables $(u,v)$, the autonomous system becomes
\begin{equation}
\frac{du}{dN}=-\gamma\,\frac{u\,v}{h(u,v)},\qquad
\frac{dv}{dN}=-3v\,\frac{v^2-1}{3v^2-1}+\varepsilon\,\frac{u}{h(u,v)\,(3v^2-1)},
\label{eq:system_uv_clean}
\end{equation}
with
\begin{equation}
h^2(u,v)=-1+\alpha\,(3v^4-2v^2)+\xi\,u.
\label{eq:h2_uv_clean}
\end{equation}

\begin{figure}[t]
\centering
\begin{tabular}{cc}
\includegraphics[width=0.47\linewidth]{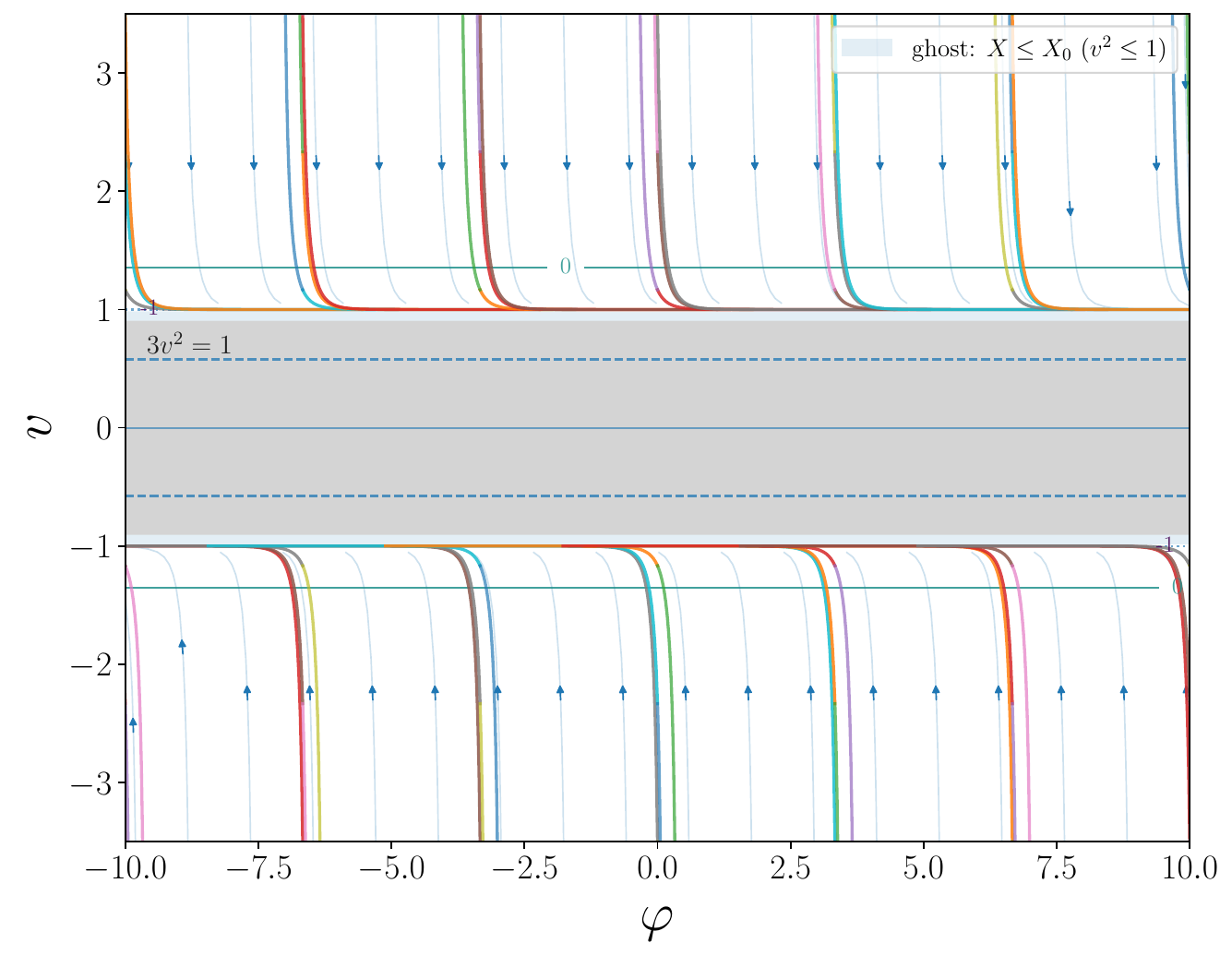} &
\includegraphics[width=0.47\linewidth]{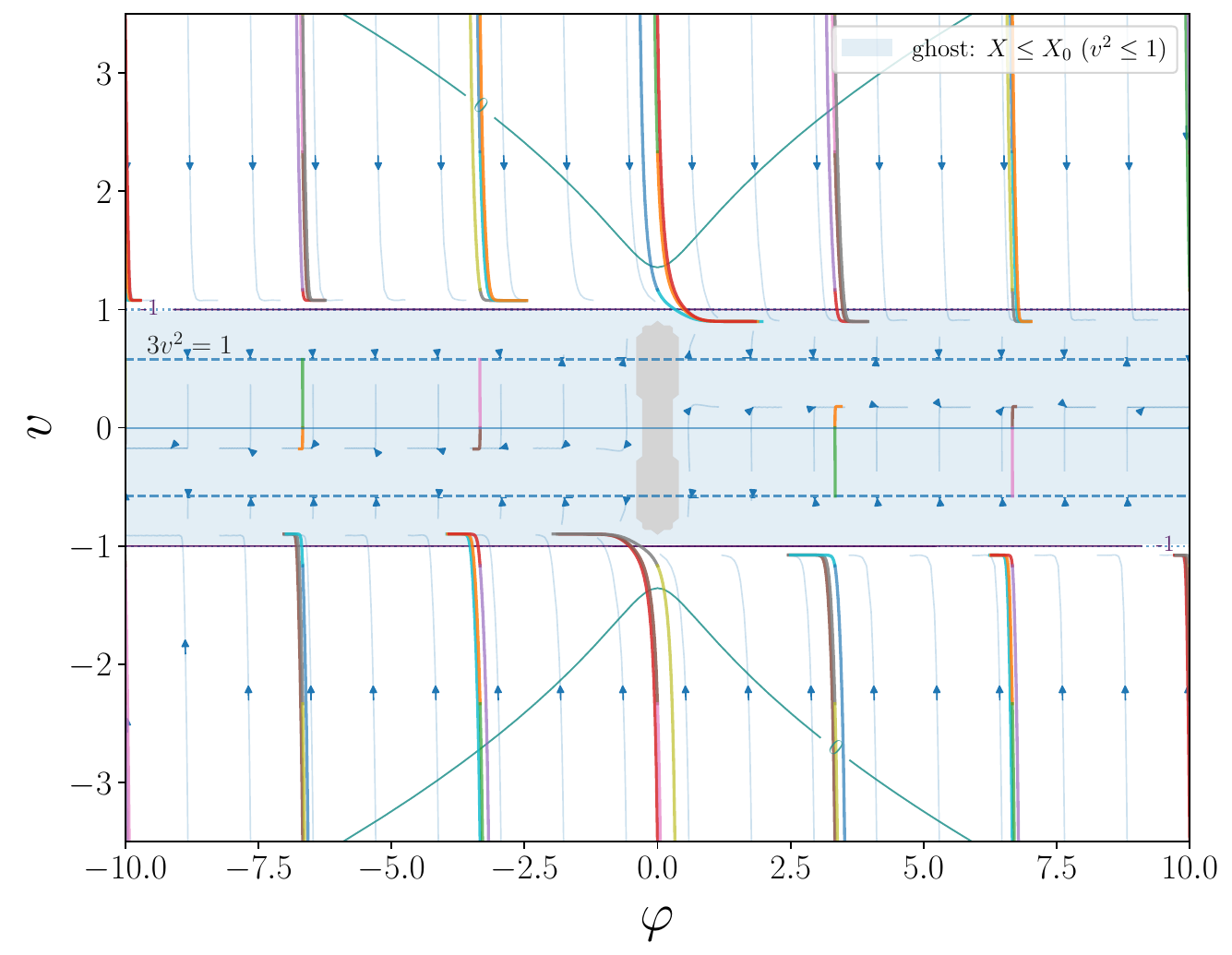} \\
\multicolumn{2}{c}{\includegraphics[width=0.47\linewidth]{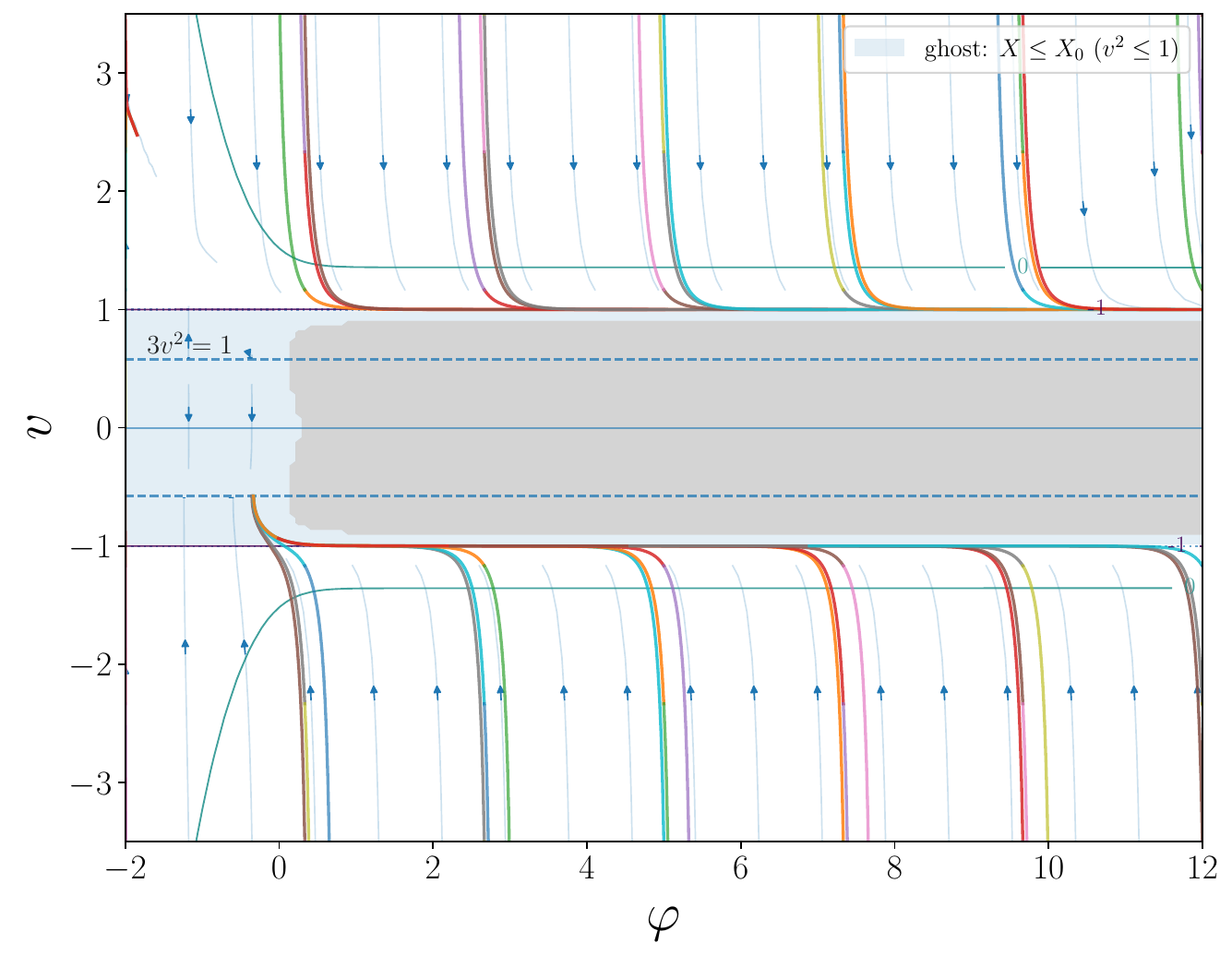}}
\end{tabular}
\caption{Late-time phase portraits in the $(\varphi,v)$ plane for (top left) the purely kinetic sector $V=0$, (top right) the quadratic potential, and (bottom) the exponential potential. The light-blue shaded band indicates the ghost domain $v^2\le 1$ (i.e.\ $X\le X_0$), excluded by the no-ghost condition $F_X>0$ (equivalently $X>X_0$). The gray shaded region corresponds to $h^2(\varphi,v)\le 0$, where the Friedmann equation would give an imaginary expansion rate and the phase-space flow is not physically defined. The dashed horizontal lines mark the kinetic singularity $3v^2=1$, i.e.\ $v=\pm 1/\sqrt{3}$, where the evolution equations become singular and the flow splits into disconnected branches.}
\label{fig:late_portraits_varphi_v}
\end{figure}

\begin{figure}[t]
\centering
\includegraphics[width=0.55\linewidth]{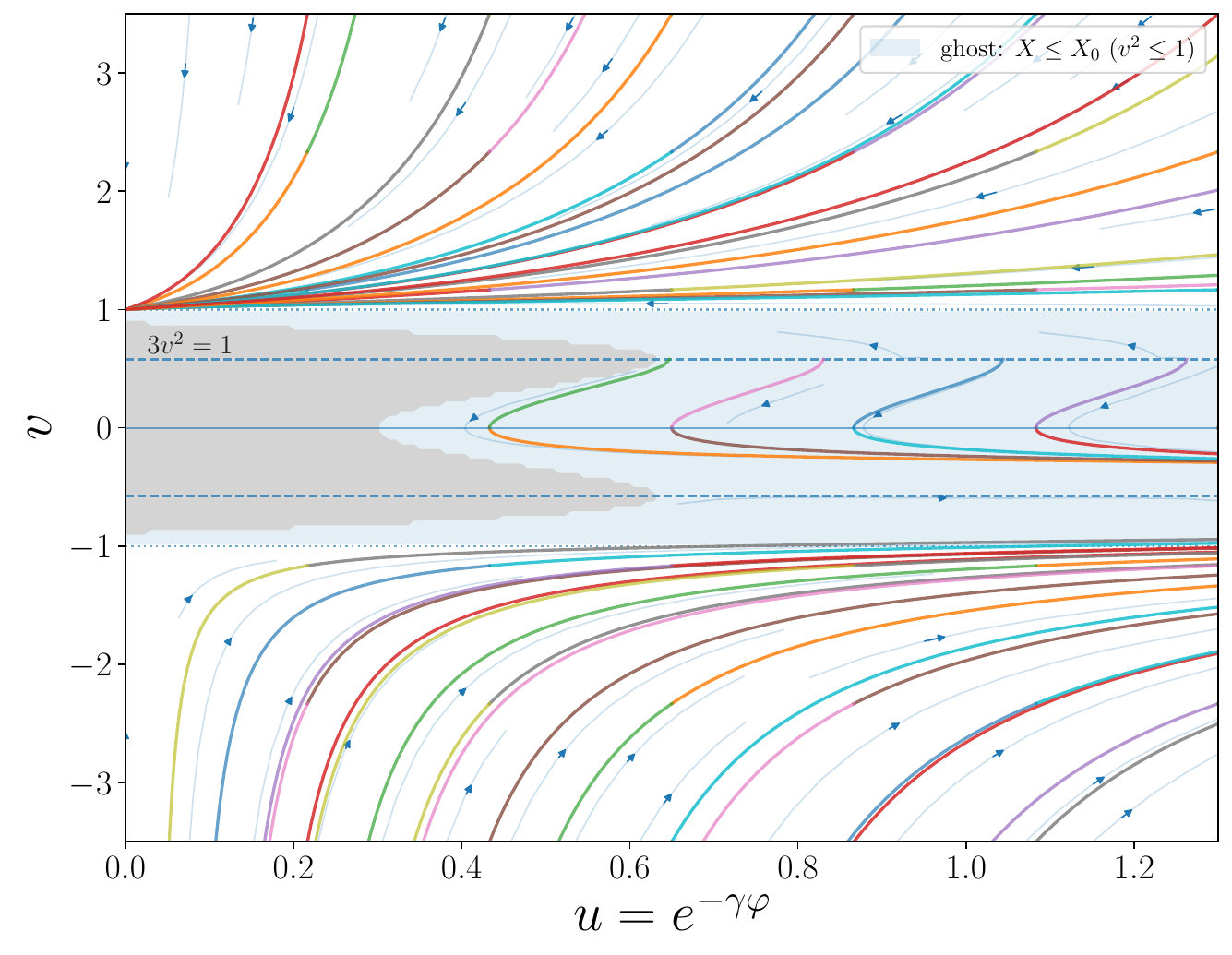}
\caption{Compactified late-time phase portrait for the exponential potential in the $(u,v)$ plane, with $u=e^{-\gamma\varphi}$. The light-blue shaded band indicates the ghost domain $v^2\le 1$, while the gray shaded region corresponds to $h^2(u,v)\le 0$ (no real $H$). The dashed horizontal lines mark the kinetic singularity $v=\pm 1/\sqrt{3}$. }
\label{fig:late_portraits_u_v}
\end{figure}

The phase portraits corresponding to \eqref{eq:system_general_clean}--\eqref{eq:h2_exp_clean} are shown in Fig.~\ref{fig:late_portraits_varphi_v}. The shaded region indicates the ghost domain $v^2\le 1$, which is excluded by the no-ghost condition $F_X>0$ (equivalently $X>X_0$). The dashed horizontal lines mark the kinetic singularity $3v^2=1$, i.e.\ $v=\pm 1/\sqrt{3}$, where the evolution equations become singular and trajectories are separated into disconnected branches. We ilustrated the compactified portrait  in Fig.~\ref{fig:late_portraits_u_v}.

For the exponential potential, the lines $v=\pm 1$ act as one-way barriers, but with different orientations. Indeed, evaluating \eqref{eq:system_general_clean} at $v\pm1$  one finds that the kinetic contribution vanishes and the potential source fixes the sign of the flow, yielding $dv/dN>0$ on both boundaries. Hence, $v=1$ repels trajectories on the $v>1$ branch, preventing them from crossing into the ghost region. By contrast, $v=-1$ attract trajectories from the $v<-1$ side toward increasing $v$, so they can cross into $v^2 \le 1$; our numerical exploration shows that such solutions may evolve into the  ghost region. In what our parameter explorations in the following section we  restrict to the sector free of ghosts $v>1$ by choosing large positive values for $\dot \phi$ at the initial conditions of our code.

For the quadratic potential, the extra force term proportional to $\delta$ in \eqref{eq:h2_quad_clean} can, for sufficiently large effective mass, drive some trajectories toward the boundary $v^2=1$ and into the ghost region. The portraits shown here are intended to illustrate this qualitative effect. In the cosmologically relevant parameter space, however, the physical prior \eqref{mphi_limit} enforces an ultralight $m_\phi$, which suppresses the $\delta$ term and keeps the dynamics very close to the purely kinetic flow  preserving the ghost-free condition. However the sign of $dv/dN$ (negative for for $v= 1$, and positive for $v= -1$) indicates that the Universe will inevitably enter the  ghost  region in  the future of its evolution.

In terms of the dynamical variables, the propagation speed of scalar fluctuations is
\begin{equation}
c_s^2
=\frac{P_X}{\rho_X}=\frac{v^2-1}{3v^2-1}\,,
\label{eq:cs2_v}
\end{equation}
The sign change of $c_s^2$ is tied to the same condition that triggers the ghost sector: it can only occur when $X<X_0$ (equivalently $v^2<1$). Therefore, all trajectories that remain in the ghost-free branch $v>1$ automatically satisfy $0<c_s^2<1/3$ and are free of gradient instabilities.

\section{Data and Methodology}
\label{sec:data}

This section describes the observational datasets and numerical tools employed to test the K-essence models introduced in Section~\ref{sec:k-essence}. We aim to assess the viability of both purely kinetic and potential-extended scenarios by confronting them with current cosmological observations spanning early and late times.

Observations include baryon acoustic oscillation (BAO) measurements from the first data release of the Dark Energy Spectroscopic Instrument (DESI 
DR1)~\cite{DESI:2024mwx}. These measurements constrain the angular diameter 
distance $D_A(z)$ and the Hubble distance $D_H(z)$ across seven redshift bins 
extending to $z \sim 1.7$. The dataset includes the full covariance matrix between $D_A$ and $D_H$, providing robust sensitivity to the late-time expansion 
history and the shape of the matter power spectrum.

For early Universe constraints, we use the \textit{Planck} 2018 legacy release (PR3), incorporating temperature and polarization anisotropies (TT, TE, EE), as well as the CMB lensing potential power spectrum~\cite{Planck:2018vyg}.
  
In all K-essence scenarios, the scalar field behaves as a radiation-like fluid during the early Universe, contributing non-negligibly to the total energy density before the matter-radiation equality and thereby affecting the inferred value of $N_{\mathrm{eff}}$ and therefore the nucleosynthesis processes.

To constrain the radiation content at early times, we include Big Bang Nucleosynthesis (BBN) likelihoods in our analysis, making use of both the primordial helium abundance $Y_p$ and the primordial deuterium fraction D/H.  
In practice, we adopt the implementation provided in \texttt{MontePython}, which combines measurements of deuterium from high-redshift quasar absorption systems~\cite{Cooke:2018qzw}, helium from recombination lines~\cite{Aver:2015iza}, and helium from emission line measurements in metal-poor $\rm{H}_2$ regions~\cite{Peimbert:2016bdg}.  
This combined dataset implies the conservative upper bound $N_{\mathrm{eff}} < 3.15$ at 95\% C.L., as summarized in Ref.~\cite{Pitrou:2018cgg}, which we adopt as a prior in our analysis.

Theoretical predictions for the CMB power spectra and the linear matter power spectrum are computed with \texttt{Hi\_CLASS}~\cite{Blas:2011rf, Zumalacarregui:2016pph, Bellini:2019syt}. 
Where the matter power spectra $P(k)$ is defined in terms of the matter density contrast $\delta_{m}$ by
\begin{equation}
    \left< \delta_{m}(\bold{k}) \delta_{m}(\bold{k}')  \right>
    =
    (2\pi)^3 \delta^{(3)}(\bold{k}-\bold{k}')P(k)\,.
    \label{mpk}
\end{equation}
In our analysis we work within the shift-symmetric subclass already supported by \texttt{Hi\_CLASS}, $G_2(X,\phi)=F(X)-V(\phi)$, and implement specifically the Scherrer kinetic function together with the quadratic and exponential potentials. 
\texttt{Hi\_CLASS} is an extended version of the Boltzmann code \texttt{CLASS} that solves the background and linear perturbation equations for general Horndeski and K-essence models; we simply provide the explicit forms of $F(X)$ and $V(\phi)$ corresponding to the scenarios studied here as well as a better suited first point to the shooting method  used to find the initial conditions of the field Eq. \eqref{shooting}.
\footnote{Our modified code can be found in \url{https://github.com/Eladio-Moreno/k-essence-dynamics}.}
These additions allow us to compute the CMB and matter power spectra required for the likelihood analysis without altering the underlying \texttt{Hi\_CLASS} structure.
In our analysis, we fix the effective number of relativistic neutrino species to $N_{\mathrm{eff,\nu}} = 3.046$ while the contribution to the relativistic degrees of freedom varies with time and
with the initial conditions of the field.

An important aspect of the quadratic potential implementation is the choice of an ultralight scalar field mass. This follows from basic order of magnitude arguments. If we assume that the field starts at the minimum of its potential at early times, $V(t=0)=0$, which is equivalent to setting $\phi(t=0)=0$ as an initial condition. In the regime where the kinetic term $X$ remains close to its minimum value $X_0$, which we will show it's the case to satisfy the BBN constrain, the field evolves approximately as $ \phi(t) \simeq \sqrt{2X_0} \, t$.
The potential energy at the present epoch then takes the form $ V(\phi_0) \simeq m_\phi^2 X_0 t_0^2$.
Requiring that this does not exceed the observed dark energy density, $V_{\rm max} \simeq 0.7 \rho_c \simeq 6 \times 10^{-11}~\mathrm{eV}^4$, yields an upper limit on the scalar field mass:
\begin{equation}\label{mphi_limit}
m_\phi \lesssim \left( \frac{V_{\rm max}}{X_0 t_0^2} \right)^{1/2} \simeq \frac{1.18 \times 10^{-38}~\mathrm{eV}^3}{\sqrt{X_0}},
\end{equation}
where we have used $t_0 \approx 6.57 \times 10^{32}~\mathrm{eV}^{-1}$ as the age of the Universe. This estimate is consistent with the values naturally obtained in our MCMC chains and with similar results in related quintessence models~\cite{6pqs-xjln}. In practice, we fix $m_\phi = 10^{-38}~\mathrm{eV}$, which both respects this physical prior and ensures numerical stability in \texttt{Hi\_CLASS}. This choice maintains the potential subdominant throughout cosmic history, allowing it to influence the dynamics only at very late times.

Bayesian parameter inference is performed using the Markov Chain Monte Carlo (MCMC) engine \texttt{MontePython}~\cite{Brinckmann:2018cvx, Benjamin_Audren_2013}, with our modified \texttt{Hi\_CLASS}. Beside the nuisance parameters, we sample the standard 6 $\Lambda$CDM parameters
$\omega_b$ (physical baryon density), $\omega_{\rm dm}$ (physical dark matter density that we replace by the matter contribution of the scalar field), $\theta_s$ (angular size of the sound horizon at recombination),  
$\tau$ (optical depth to reionization), $n_s$ (scalar spectral index), and $A_s$ (amplitude of primordial scalar perturbations).
We then include the additional K-essence parameters: $\theta$ for the purely kinetic case and $(\theta, V_0, \lambda)$ or $(\theta,m_\phi)$ for the exponential and quadratic potentials respectively. 
The total likelihood function is given by the product of the Planck, DESI BAO, and BBN contributions.

Posterior distributions are analyzed using \texttt{GetDist}~\cite{Lewis:2019xzd}, which provides high-precision kernel density estimation and visualization tools. 
This framework allows for a consistent exploration of how K-essence dynamics affect both the expansion history and the growth of structure, while remaining compatible with constraints on radiation content and primordial nucleosynthesis.

Finally, we emphasize that a consistent treatment of K-essence cosmologies requires evolving both the background and linear perturbations.  
In particular, the clustering properties of the scalar field can leave imprints on the matter power spectrum and CMB anisotropies, making the perturbative sector essential for testing these models against observations.  
A comprehensive overview of perturbation dynamics in K-essence and related scalar field scenarios can be found in the review by Amendola and Tsujikawa~\cite{Kase:2018aps} and~\cite{Bertacca:2011in}, which highlights their relevance for distinguishing these models from $\Lambda$CDM.

\section{Results and Discussion}
\label{sec:results_discussion}
\noindent
This section presents the main results of our analysis, including both background and linear perturbation evolution, as well as cosmological parameter constraints. We assess the viability of generalized K-essence modelsas alternatives to the \(\Lambda\)CDM paradigm in the light of current observational data, with three
different Lagrangians, the purely kinetic quadratic Lagrangian originally introduced by Scherrer in~\cite{Scherrer:2004au} and our extended versions incorporating exponential and quadratic potentials.

\subsection{Background Evolution}\label{sec:background}

\begin{figure}[H]
\centering
\includegraphics[width=0.6\linewidth]{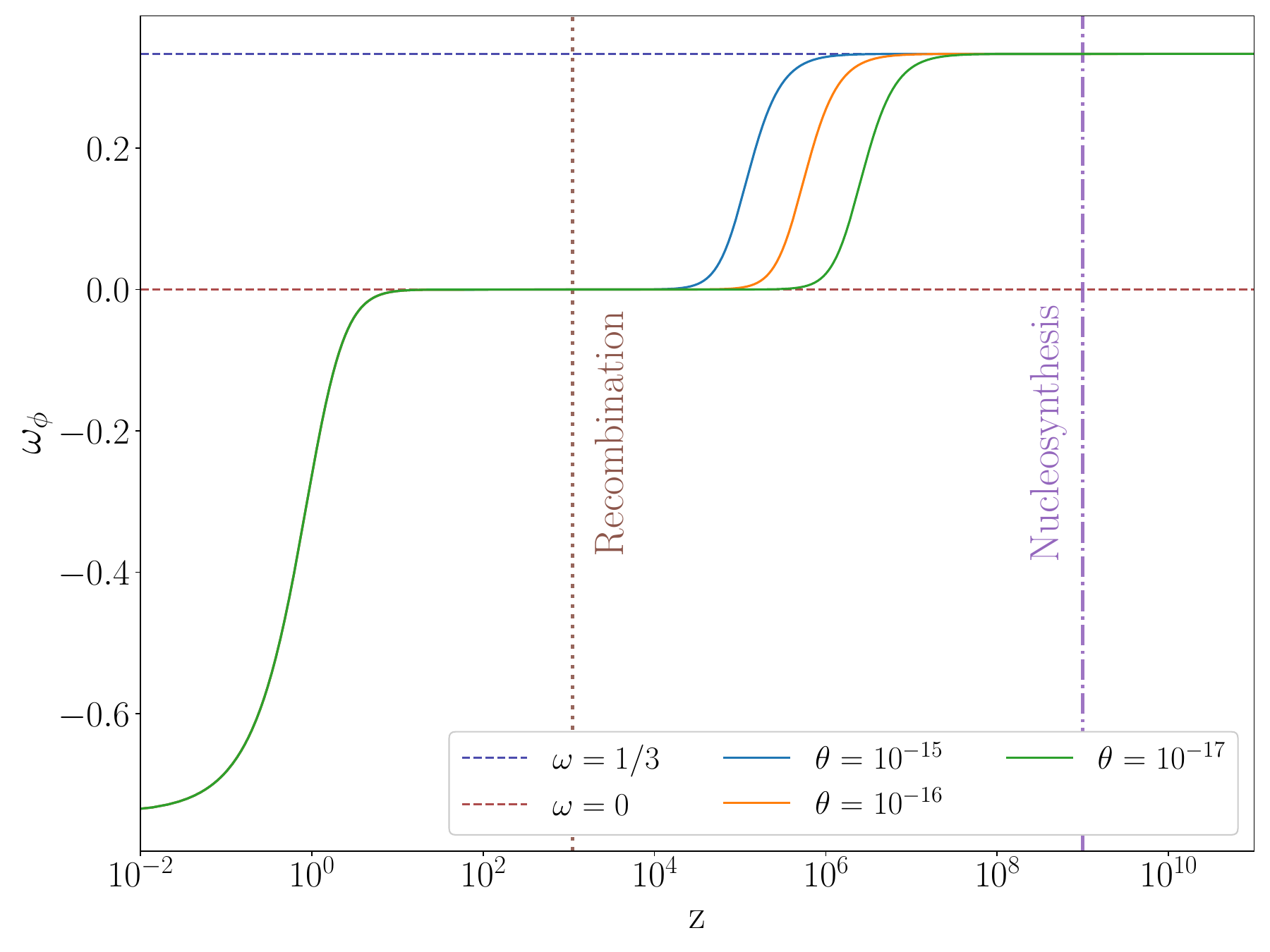}
\caption{Redshift evolution of the scalar field equation of state \(w_\phi\) for various values of \(\theta\). Vertical dashed lines indicate the epochs of BBN and recombination. We see that the field behaves as radiation at high redshifts,
transitioning to a matter-like behavior before recombination. Therefore the bounds over extra radiation components on BBN are very important con constrain the  model.}
\label{fig:cinetico_w}
\end{figure}

We begin with the purely kinetic model, which provides a minimal realization of a unified dark sector.
With respect to the $\Lambda$CDM model parameters, this model adds one degree of freedom, changing the $\Omega_{dm}$ and $\Omega_\Lambda$ to $F_0$ and $F_2$ as can be seen in \eqref{relationF2_F0}, but adding a new parammeter $\theta$ asociated with the initial conditions of the field.
Figure~\ref{fig:cinetico_w} shows the redshift evolution of the scalar field equation of state \(w_\phi\) for various values of the initial condition \(\theta\). At large redshifts, the field initially behaves as a radiation fluid (\(w_\phi \to 1/3\)) that later transitions to a matter behavior (\(w_\phi \simeq 0\)) before recombination, and eventually drives cosmic acceleration through the constant term \(F_0\), which acts effectively as a cosmological constant.

In the early Universe, the scalar field contributes as an additional relativistic species, enhancing the total radiation density. As discussed in Sec.~\ref{sec:data}, this behavior is constrained by Big Bang Nucleosynthesis (BBN), which imposes the bound $N_{\rm eff} < 3.15$ at 95\% C.L.~\cite{Cooke:2017cwo,Aver:2015iza,Peimbert:2016bdg,Pitrou:2018cgg}.
This bound leads to a $\Delta N_{\rm eff}<0.104$ for the contribution of the K-essence field to the early radiation density which
from (\ref{bound_neff}) give a bound to the initial condition parameter of
\begin{equation}\label{theta:constriction}
    \theta < 2.97\times 10^{-16}\,,
\end{equation}
for the purely kinetic model. This condition comes only from the
BBN observations and order of magnitude assumptions for
$\Omega_\gamma$ and $\Omega_{dm}$, and
is close to the condition obtained in the next
subsection combining data from Planck+DESI DR1+BBN, which gives 
$\theta \leq 1.97 \times 10^{-16}$ at 95\% C.L.
as derived from our MontePython chains. 

We stress that the small upper bound on $\theta$ in Eq.~\eqref{theta:constriction} does not imply a negligible contribution of the scalar field to the dark matter sector. The actual dark matter abundance sourced by the scalar field is fixed by $\Omega_{\rm dm}$ through Eq.~\eqref{relationF2_F0}, while a small $\theta$ is required by BBN constraints to suppress an excessive early-time radiation-like contribution. The smallness of $\theta$ is compensated by a large value off $F_2$ in equation \eqref{scherrer_density}.
\begin{figure}[H]
\centering
\includegraphics[width=0.6\linewidth]{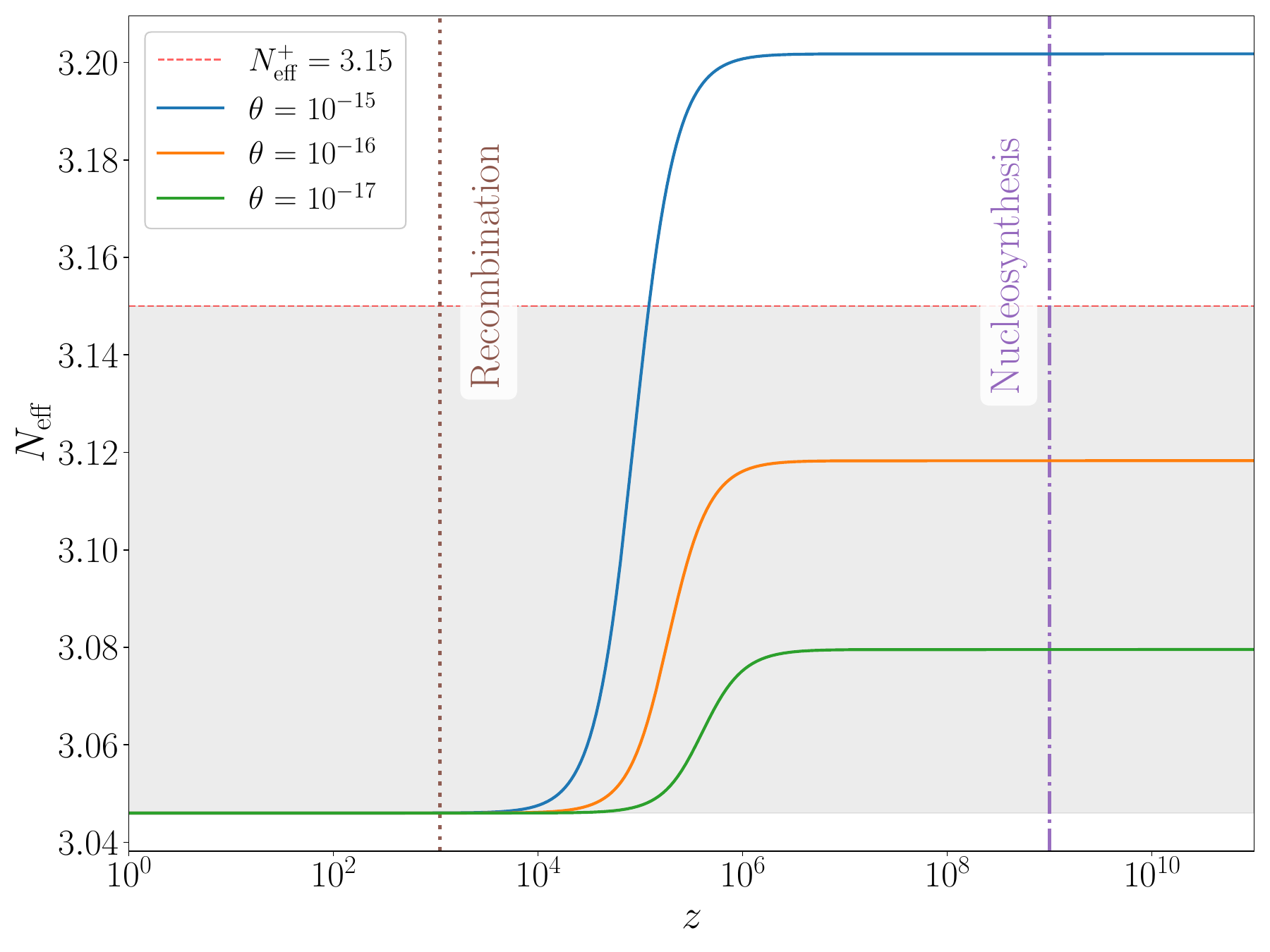}
\caption{Redshift evolution of the effective number of relativistic species \(N_{\mathrm eff}\). The gray shaded region denotes the upper bound from BBN constraints~\cite{Cooke:2017cwo,Aver:2015iza,Peimbert:2016bdg,Pitrou:2018cgg}. As the scalar field becomes non-relativistic, its contribution to the radiation density vanishes, restoring the standard value of \(N_{\mathrm eff}\).}
\label{fig:cinetico_Neff}
\end{figure}
From Eq.~\eqref{approx} evaluated at recombination
we see that the field needs to be very close to its minimum already at this epoch
\begin{equation}
    \frac{X(z_{\rm{rec}})-X_0}{X_0}\le 2.63 \times 10^{-7}.
\end{equation}
Figure~\ref{fig:cinetico_Neff} illustrates the evolution of $N_{\rm eff}$ for representative values of $\theta$. The gray band marks the range allowed by BBN, showing that the relativistic phase of the field must fade sufficiently early so that its contribution vanishes before recombination, restoring the standard value of $N_{\rm eff}$.

\begin{figure*}[t]
    \centering
    \begin{subfigure}[t]{0.49\textwidth}
        \centering
        \includegraphics[width=\linewidth]{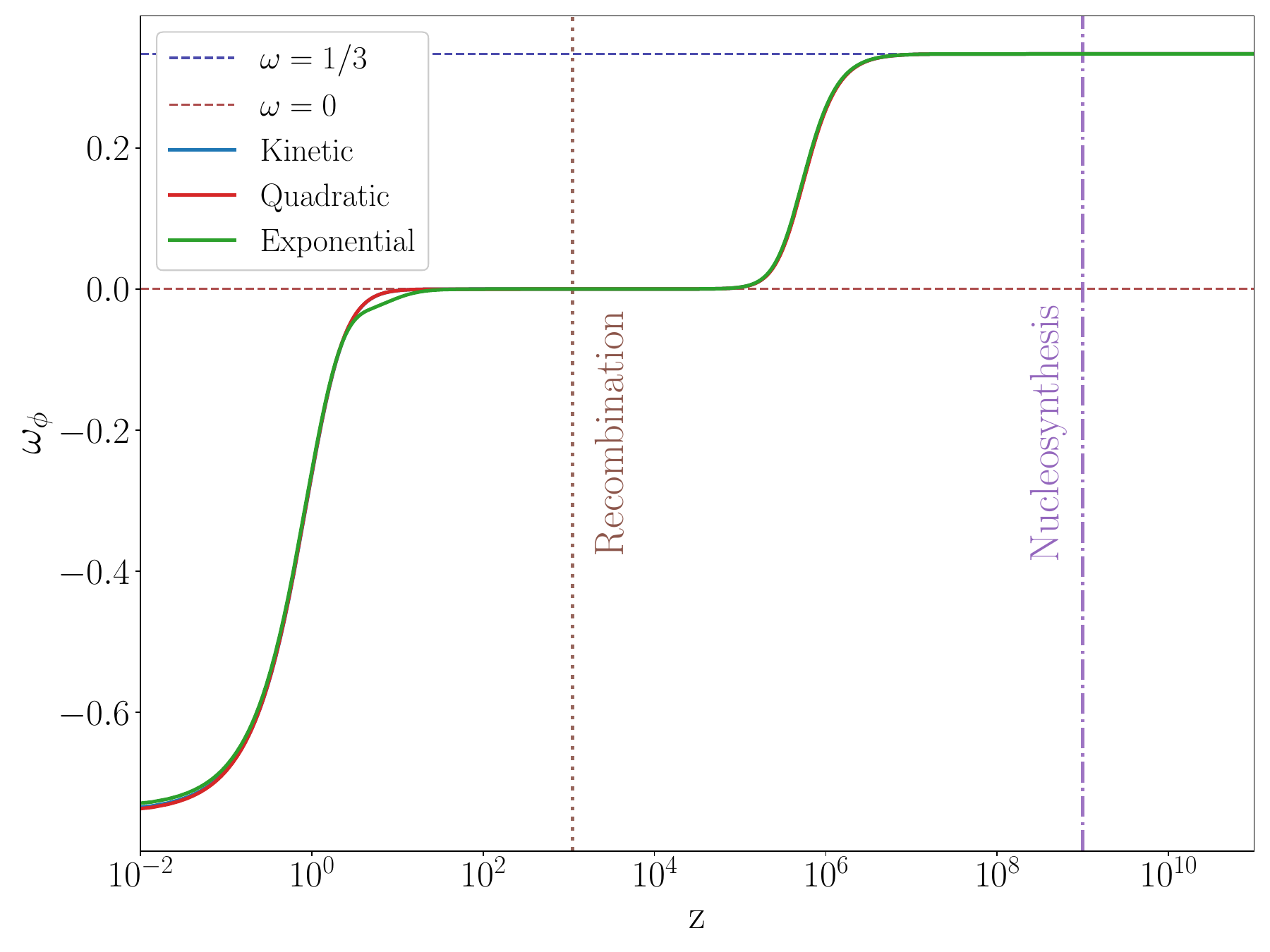}
        \caption{Equation-of-state parameter $\omega_\phi(z)$ for the three potentials considered.}
        \label{fig:wphi_compare}
    \end{subfigure}\hfill
    \begin{subfigure}[t]{0.49\textwidth}
        \centering
        \includegraphics[width=\linewidth]{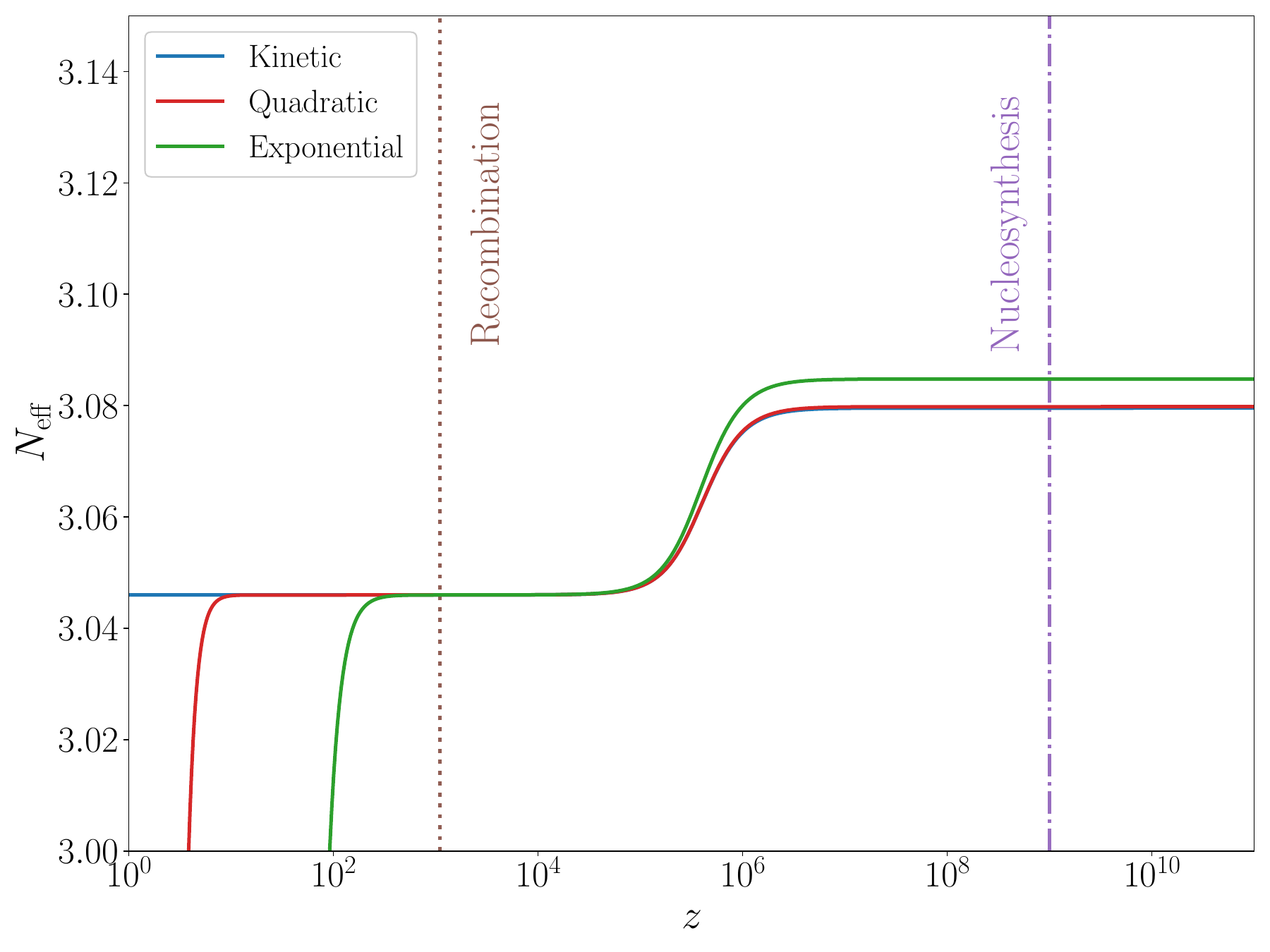}
        \caption{Effective $N_{\rm eff}(z)$ proxy for the same models.}
        \label{fig:neff_compare}
    \end{subfigure}

    \caption{Comparison of the background evolution in the unified dark sector models.
Left: equation-of-state parameter $\omega_\phi(z)$.
Right: the effective $N_{\rm eff}(z)$ diagnostic computed from the total relativistic budget in the code.
For the exponential potential we use $V_0=4.6\times 10^{-3}\,\mathrm{eV}^4$ and $\lambda=1.2\times 10^{-30}\,\mathrm{eV}^{-1}$, while for the quadratic potential we use $m_\phi=5.1\times 10^{-38}\,\mathrm{eV}$; in both cases we set $\theta=10^{-17}$.
Negative values of the plotted effective $N_{\rm eff}$ arise when the scalar pressure becomes negative and should not be taken into account as $N_{\rm eff}$ is measured during recombination and nucleosynthesis.}
    \label{fig:wphi_neff_joint}
\end{figure*}

For visual reference, we also included Figure \ref{fig:wphi_neff_joint} with plots of $\omega_\phi$ and $N_{\rm eff}$ in the presence of a potential. We find that the equation of state evolution is only marginally modified, while  the $N_{\rm eff}$ increases due to the potential at early times. $N_{\rm eff}$ goes  to negative values as the formula used to compute it, given by
\begin{equation}
    N_{\rm eff} =\frac{8}{7}
    \left( \frac{11}{4}\right)^{4/3}
    \frac{3p_\nu + 3 p_\phi}{\rho_\gamma}
\end{equation}
is only suited for high redshifts where this parameter is measured.

\subsection{Perturbations}
\label{sec:perturbations}

The behavior of linear perturbations in K-essence models reveals distinct signatures in the matter power spectrum $P(k)$, driven by the background dynamics of each scenario. Throughout this section we fix $X_0 = (0.12~\mathrm{eV})^4$. This choice ensures direct comparability between the models and avoids introducing additional parameter degeneracies as $X_0$ is very poorly constrained
by the data. In the purely kinetic scenario, the qualitative shape of $P(k)$ is insensitive to the specific choice of $X_0$, whereas in models with scalar potentials the variation of $X_0$ can affect the small-scale behavior. The latter case will be further discussed in the parameter constraints analysis.

\begin{figure*}[ht]
\centering
\begin{subfigure}[b]{0.32\textwidth}
\includegraphics[width=\textwidth]{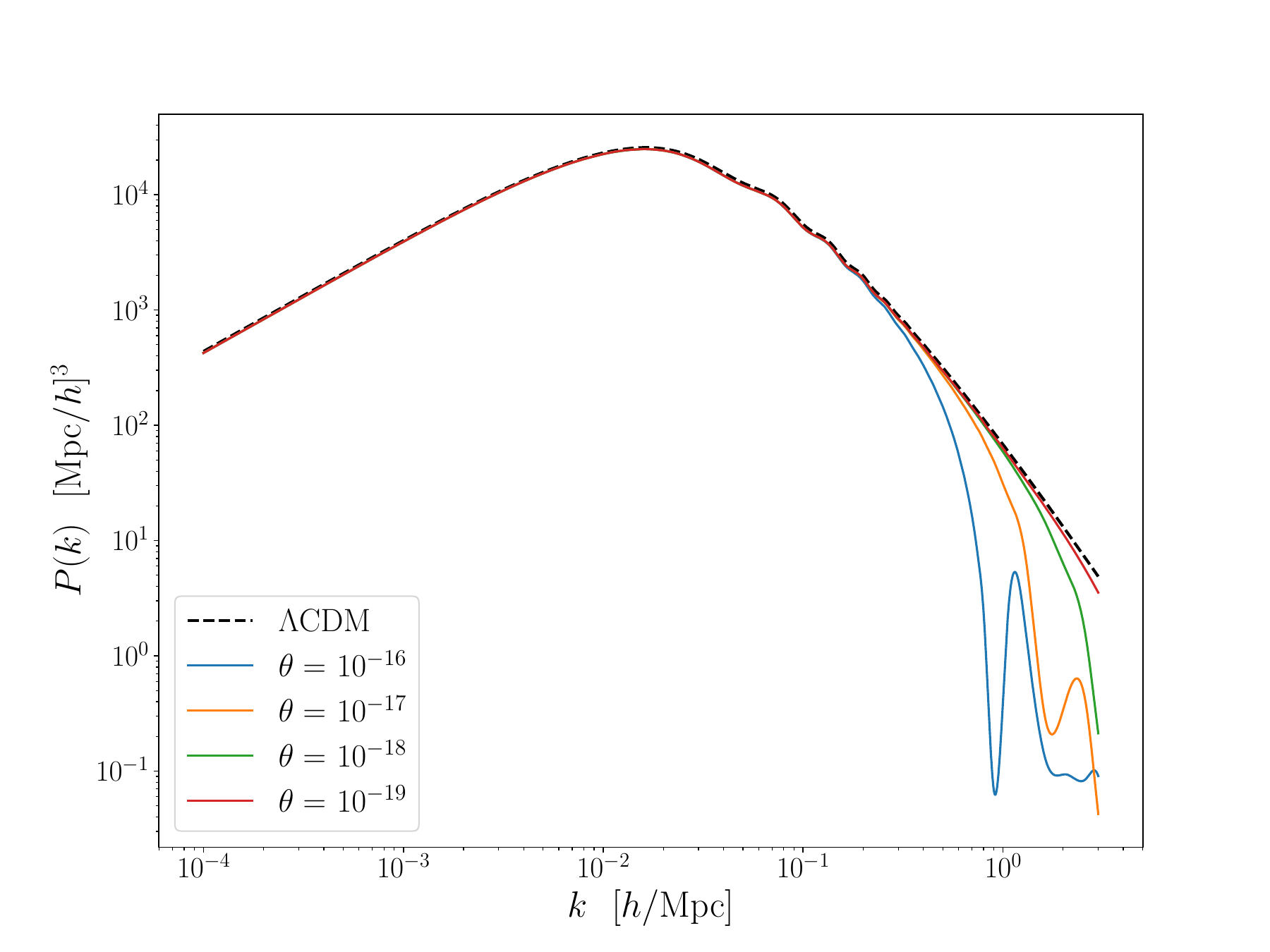}
\caption{Purely kinetic model}
\label{fig:cinetico_matter}
\end{subfigure}
\begin{subfigure}[b]{0.32\textwidth}
\includegraphics[width=\textwidth]{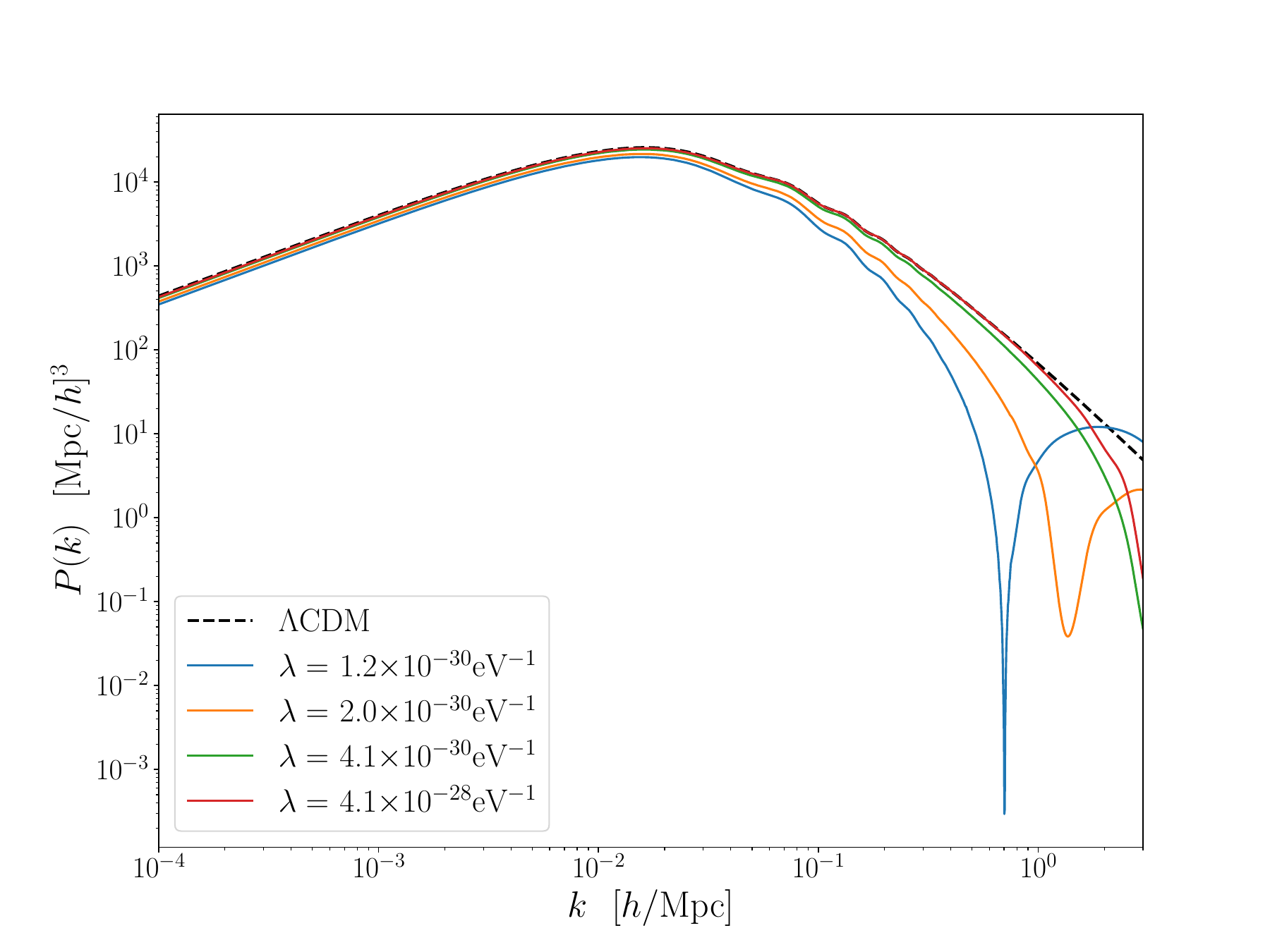}
\caption{Exponential potential}
\label{fig:exponencial_matter}
\end{subfigure}
\begin{subfigure}[b]{0.32\textwidth}
\includegraphics[width=\textwidth]{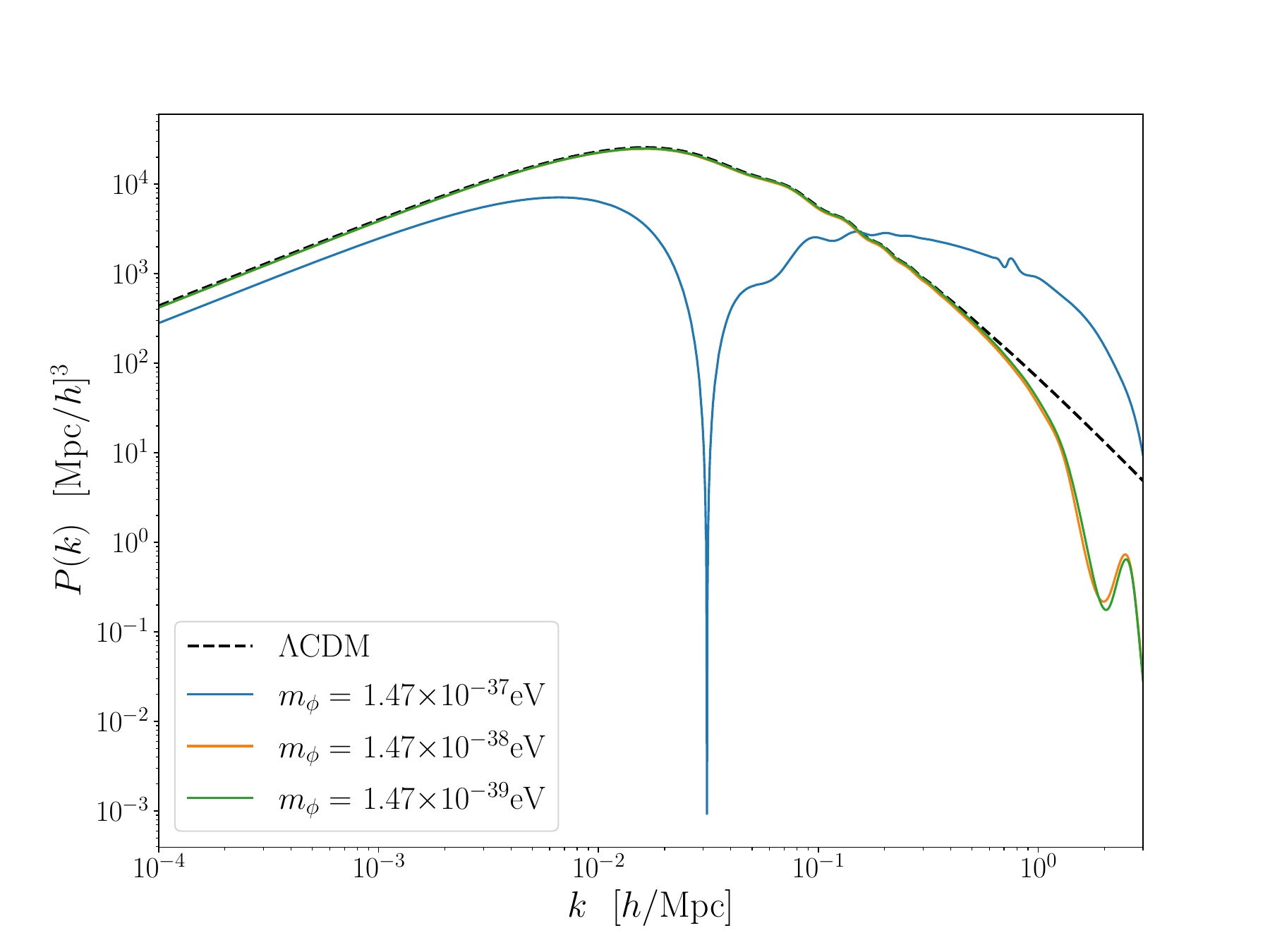}
\caption{Quadratic potential}
\label{fig:quadratic_matter}
\end{subfigure}
\caption{Linear matter power spectrum \eqref{mpk} at $z=0$ for the three K-essence scenarios.  
\textbf{(a)} Purely kinetic model: larger $\theta$ values delay the radiation-to-matter transition and suppressing small scale perturbations (high $k$ values), while smaller $\theta$'s recover the $\Lambda$CDM shape.  
\textbf{(b)} Exponential potential: small $\lambda$ values 
suppress the perturbations at small scales for a wider range of $k$'s,
while large $\lambda$ recover the kinetic limit. The results are shown for $\theta = 10^{-7}$ and $V_0 = 10^{-3}\,\mathrm{eV}^4$.  
\textbf{(c)} Quadratic potential: large scalar masses induce oscillations and strong suppression, while ultralight masses $m_\phi \le 10^{-38}\rm{eV}$ approach to the purely kinetic behavior. Here we fixed $\theta = 10^{-7}$. 
For clarity, we adopt exaggerated parameter values in this figure in order to make the qualitative differences between the models more evident.
}
\label{fig:pk_allmodels}
\end{figure*}

Figure~\ref{fig:pk_allmodels} summarizes the linear matter power spectra \eqref{mpk} at $z=0$ for the three K-essence scenarios. The three cases present a suppression in
the power spectra above a certain $k$, corresponding to small scale perturbations.
Panel (a) shows the purely kinetic case, where varying the initial condition parameter $\theta$ modulates the scale of the suppression: larger $\theta$ values delay the transition from radiation-like to matter-like behavior as can be seen in figure \ref{fig:cinetico_w}, suppressing the perturbations on small scales. Conversely, smaller $\theta$ moves the transition to earlier times, recovering the $\Lambda$CDM shape at high $k$'s. This cutoff resembles that found in fuzzy dark matter models, but here it originates purely from the kinetic sector, without invoking scalar masses. A similar effect occurs on fuzzy dark matter~\cite{Hu2000,Hui2017} and ultralight scalar field models~\cite{PhysRevD.63.063506,Matos2000} but in that case it is driven by the
Compton wavelength of the field.

Panel (b) corresponds to the exponential potential model $V(\phi) = V_0 e^{-\lambda \phi}$, where the slope $\lambda$ controls the late-time dynamics. Small $\lambda$ values enhance the effect of the potential, producing stronger suppression at intermediate and small scales. As $\lambda$ increases, the potential becomes negligible and the evolution approaches the kinetic limit, recovered in the limit $\lambda \to \infty$.

Panel (c) illustrates the quadratic potential $V(\phi) = \tfrac{1}{2} m_\phi^2 \phi^2$, where the scalar mass $m_\phi$ strongly impacts perturbation evolution. For $m_\phi \le 10^{-37}~\mathrm{eV}$, rapid oscillations appear in $P(k)$, producing a sharp suppression of the perturbations on small scales. Masses below $10^{-38}~\mathrm{eV}$ yield spectra nearly indistinguishable from the purely kinetic case, as the potential remains subdominant until very late times. This behavior is consistent with the bound derived in Section~\ref{sec:data}.

\begin{figure}[H]                                                                                                                                                                                                                                                                                                                                                                                                                                                                                                                                                                                                                                                                                                                                                                                                                                                                                                                                                                                                                                                                                                                                                                                                                                                                                                                                                                                                                                                                                                                                                                                                                                                                                                                                                                                                                                                                                                                                                                                                                                                                                                                                                                                                                                                                                                                                                                                                                                                                                                                                                                                                                                                                                                                                                                                                                                                                                                                                                                                                                                                                                                                                                                                                                                                                                                                                                                                                                                                                                                                                                                                                                                                                                                                                                                                                                                                                                                                                                                                                                                                                                                                                                                                                                                                                                                                                                                                                                                                                                                                                                                                                                                                                                                                                                                                                                                                                                                                                                                                                                                                                                                                                                                                                                                                                                                                                                                                                                                                                                                                                                                                                                                                                                                                                                                                                                                                                                                                                                                                                                                                                                                                                                                                                                                                                                                                                                                                                                                                                                                                                                                                                                                                                                                                                                                                                                                                                                                                                                                                                                                                                                                                                                                                                                                                                                                                                                                                                                                                                                                                                                                                                                                                                                                                                                                                                                                                                                                                                                                                                                                                                                                                                                                                                                                                                                                                                                                                                                                                                                                                                                                                                                                                                                                                                                                                                                                                                                                                                                                                                                                                                                                                                                                                                                                                                                                                                                                                                                                                                                                                                                                                                                                                                                                                                                                                                                                                                                                                                                                                                                                                                                                                                                                                                                                                                                                                                           
\centering
\includegraphics[width=0.6\linewidth]{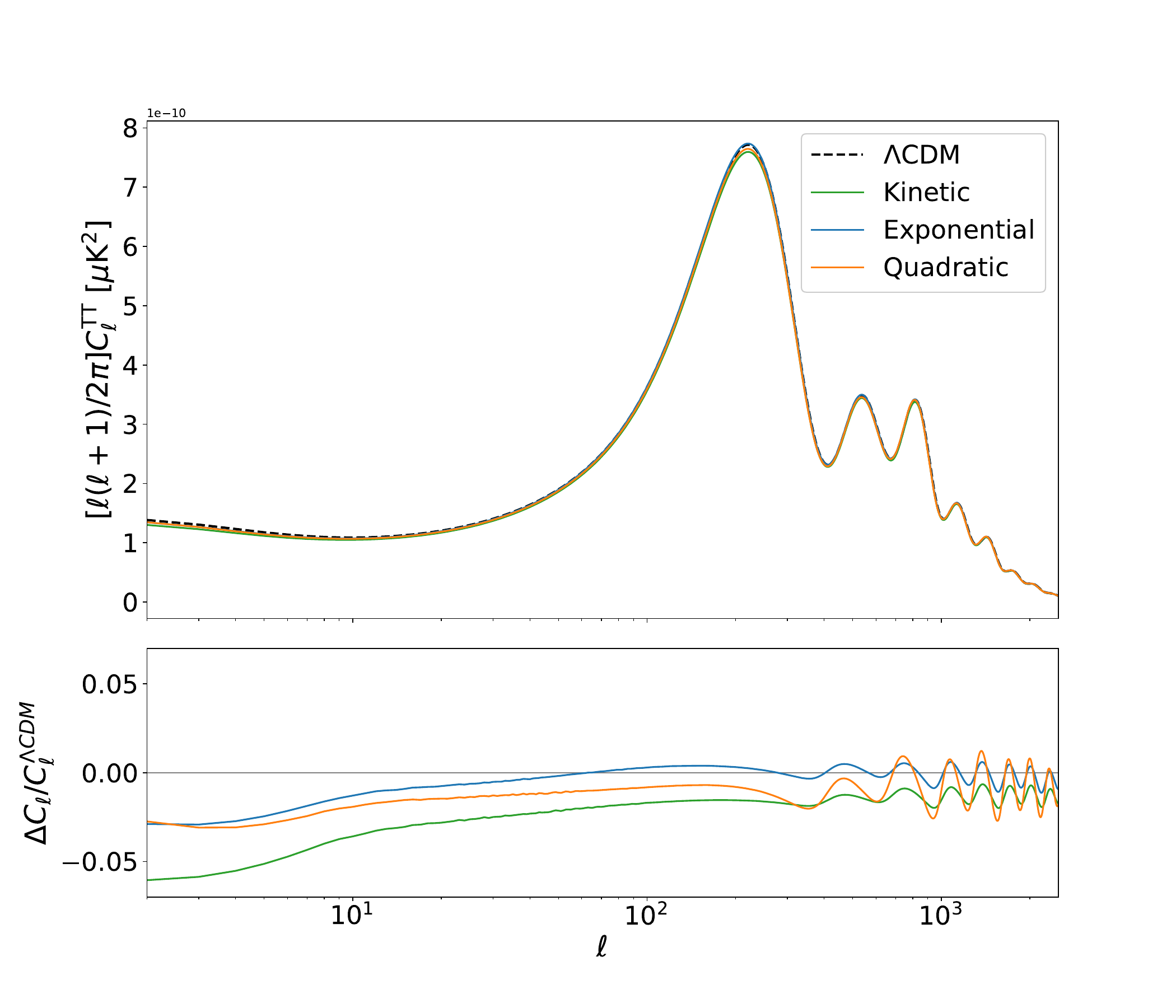}
\caption{CMB temperature power spectrum $C_\ell^{TT}$ (upper panel) and relative deviation from $\Lambda$CDM (lower panel). All K-essence models agree with \textit{Planck} data across all multipoles, with sub-percent differences at high $\ell$. The curves are obtained using the mean values of the cosmological and model-specific parameters reported in Table~\ref{table:kessence_cosmopars}.}
\label{fig:PSCMB_models}
\end{figure}

On large scales, all models reproduce the same structure as $\Lambda$CDM, ensuring compatibility with current observations. Differences arise mainly at small scales, where the transition dynamics and potential contributions modulate the growth of perturbations, leading to a characteristic suppression in $P(k)$. These deviations, while modest at present sensitivity, represent potential observational signatures for next-generation surveys.  
In particular, forthcoming galaxy surveys such as \textit{Euclid}~\cite{refId0}, the Vera C. Rubin Observatory’s Legacy Survey of Space and Time (LSST)~\cite{Ivezić_2019}, and the Dark Energy Spectroscopic Instrument (DESI DR2)~\cite{tr6y-kpc6}, together with upcoming CMB experiments like CMB-S4~\cite{Abazajian:2019eic} and the Simons Observatory~\cite{Ade_2019}, will achieve the precision needed to probe the characteristic suppressions in $P(k)$ predicted by K-essence scenarios. 
Finally, figure~\ref{fig:PSCMB_models} presents the predicted CMB temperature power spectra for all K-essence scenarios, alongside $\Lambda$CDM which presents
smaller deviations than for the matter power spectrum, but relevant due to the
high precision at which the Planck team has measured the CMB power spectrum.

\subsection{Cosmological Parameter Constraints}
\label{sec:constraints}

\begin{figure*}[ht]
\centering
\begin{subfigure}[b]{0.45\textwidth}
\includegraphics[width=\textwidth]{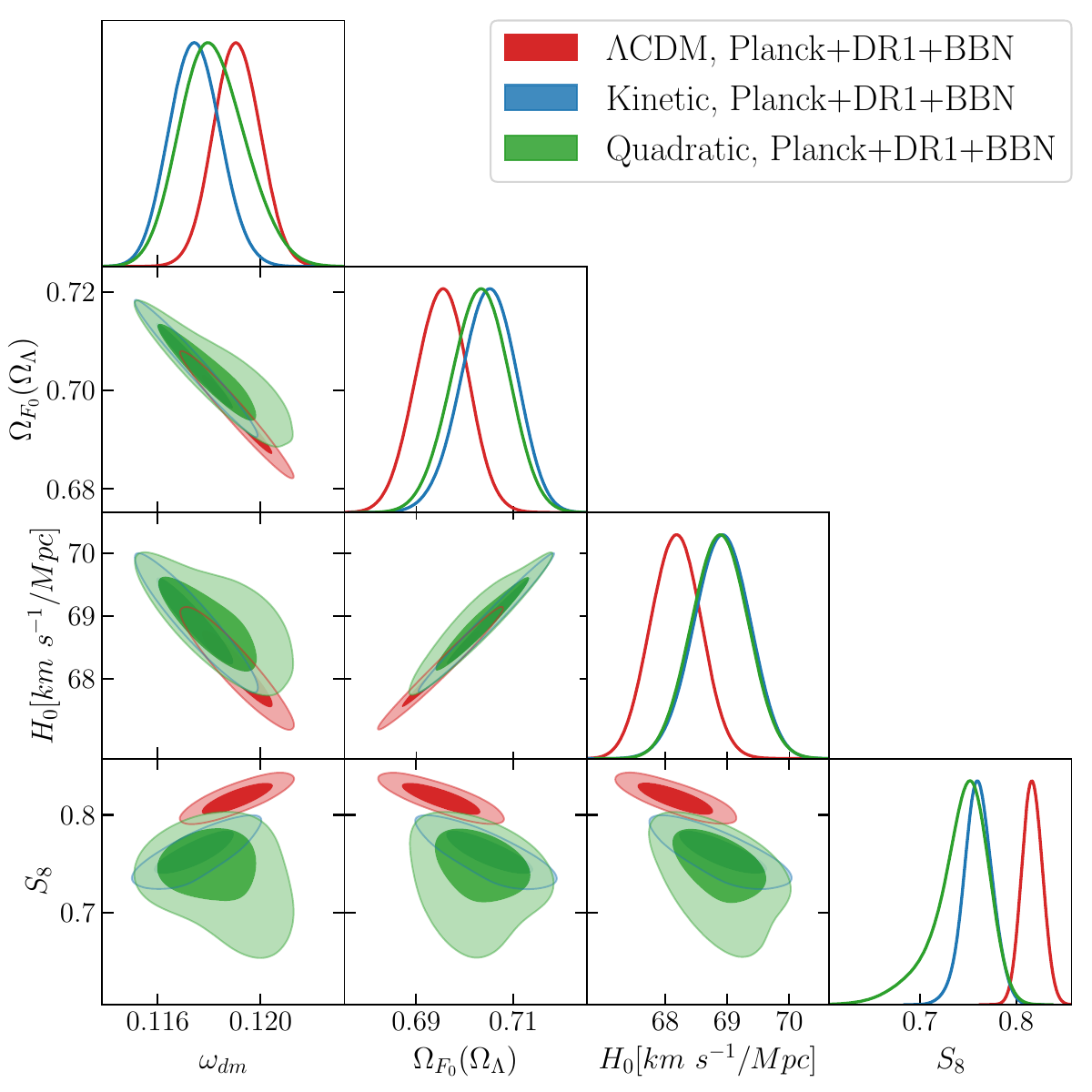}
\caption{Quadratic potential}
\label{fig:cuadratico_x01}
\end{subfigure}
\begin{subfigure}[b]{0.45\textwidth}
\includegraphics[width=\textwidth]{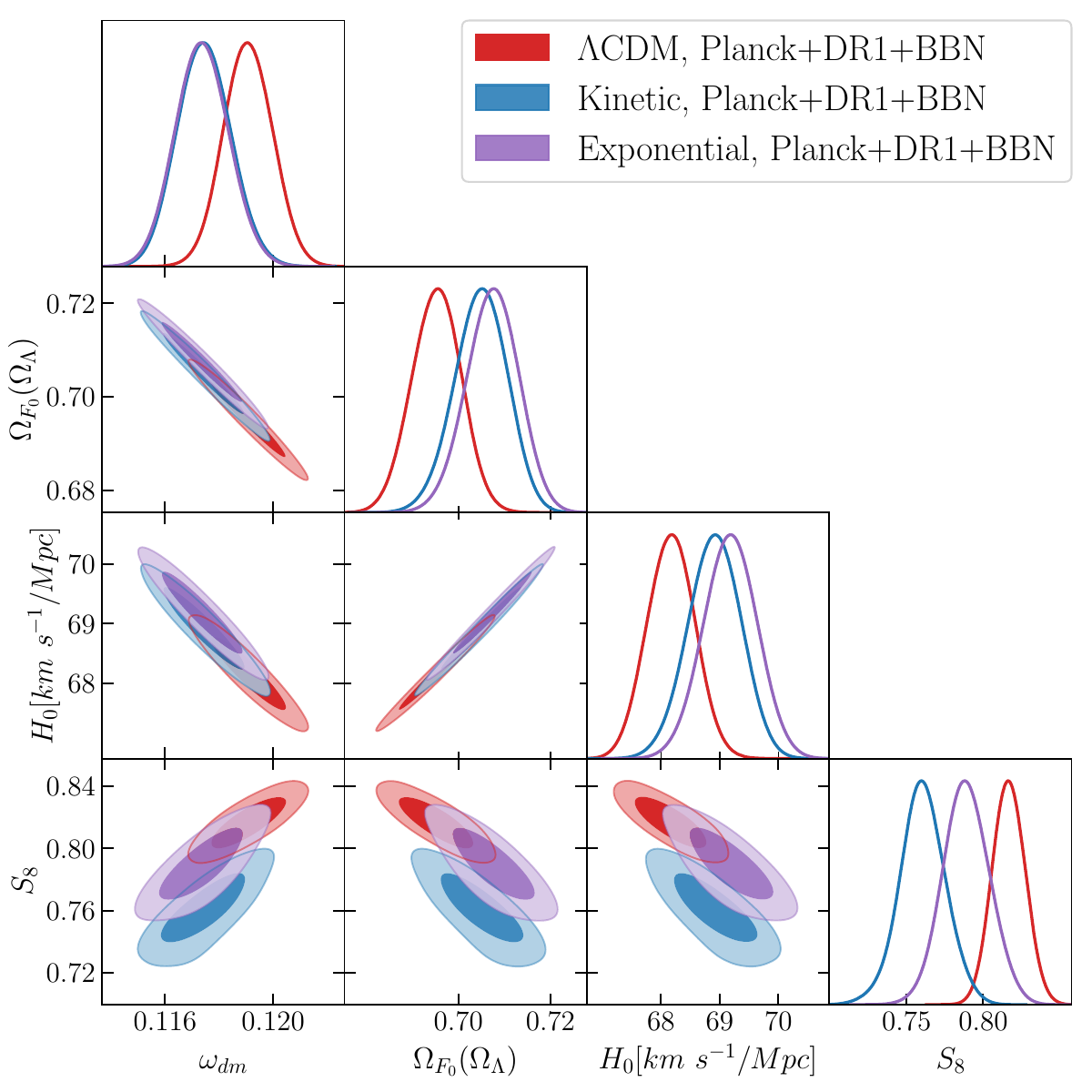}
\caption{Exponential potential}
\label{fig:exponencial_x01}
\end{subfigure}
\caption{Two-dimensional marginalized posterior distributions for 
$\omega_{\rm dm}$, $\Omega_{F_0}$, $H_0$, and $S_8$ in the three K-essence models, compared to $\Lambda$CDM. We see a shift towards
larger
$H_0$ values
reducing the tension with the SH0ES measurement.
The contours correspond to 68\% and 95\% confidence levels. }

\label{fig:posteriores_modelos_x01}
\end{figure*}

We now present the cosmological parameter constraints for the K-essence scenarios, obtained from a joint analysis of Planck 2018, DESI DR1 BAO, and BBN data. 
Figure~\ref{fig:posteriores_modelos_x01} displays the two-dimensional marginalized constraints on the parameters $\omega_{\rm dm}$, $\Omega_{F_0}$, $H_0$, and $S_8$ for the purely kinetic, quadratic, and exponential K-essence models, compared to $\Lambda$CDM.
 All three scenarios are consistent with current observations and predict coherent shifts toward higher $H_0$, partially alleviating the tension with SH0ES from $4.37\sigma$ in $\Lambda$CDM to $3.66\sigma$ (kinetic), $3.68\sigma$ (quadratic), and $3.41\sigma$ (exponential), as  summarized in table
\ref{table:kessence_cosmopars}. In the quadratic model, the scalar mass $m_\phi$ is subject to the energetic bound of Eq.~\eqref{mphi_limit}, obtained by requiring the potential energy to remain subdominant until the present epoch.  
Using $X_0 = (0.12~\mathrm{eV})^4$, this yields $m_\phi \lesssim 10^{-37}$~eV, which we adopt as a prior in the parameter estimation.  
Within this physically motivated range, the posterior distributions are nearly indistinguishable from those of the purely kinetic case, indicating that the quadratic potential does not lead to novel phenomenology under current observational constraints.

\begin{table*}[ht]
    \centering
    
    \tabcolsep=0.3cm
    \begin{tabular}{@{} llcccc @{}} 
        \hline
        Model & Kinetic & Quadratic & Exponential & $\Lambda$CDM \\
        \hline
		$\omega_{\rm dm}$($\omega_{\rm cdm}$) & $0.1174^{+0.0020}_{-0.0018}$ & $0.1182^{+0.0026}_{-0.0024}$ & $0.1174\pm 0.0019$ & $0.1191\pm0.0017$ \\
        $\Omega_{F_{0}}$($\Omega_\Lambda$) & $0.705 \pm0.011$ & $0.703^{+0.011}_{-0.012}$ & $0.707\pm 0.011$ & $0.6952^{+0.0099}_{-0.010}$ \\
        $H_0$ [km\ s$^{-1}$/Mpc] & $68.91^{+0.83}_{-0.90}$ & $68.88\pm0.87$ & $69.18^{+0.86}_{-0.80}$ & $68.17^{+0.77}_{-0.78}$ \\
        $S_8$ & $0.761^{+0.030}_{-0.031}$ & $0.744^{+0.050}_{-0.064}$ & $0.789^{+0.030}_{-0.029}$ & $0.817\pm 0.020$ \\
        10$^{17}\theta$ & $< 19.7 $ & $< 14.5$ & $< 17.8$ & -- \\
        10$^{38}m_\phi$ [eV] & -- & $< 3.48$ & -- & -- \\
        $10^3V_0$ [eV$^4$]& -- & -- & $< 2.49$ & -- \\
        $10^{27}\lambda$ [eV$^{-1}$] & -- & -- & $< 4.53$ & -- \\
        
        \hline\hline
        $\chi^2_{\rm{min}}$  & 2794.46 & 2794.04 & 2794.46 & 2799.44 \\
        Hubble Tension & 3.66$\sigma$ & 3.68$\sigma$ & 3.41$\sigma$ & 4.37$\sigma$ \\
        Extra degrees of freedom &1&2&3&---\\
        $\Delta$AIC & -2.98 & -1.40 & -0.48 & 0.0 \\
        \hline
    \end{tabular}
    \caption{Summary of the posterior mean values and $1\sigma$ uncertainties for key cosmological parameters in the Kinetic, Quadratic, and Exponential K-essence models, along with the $\Lambda$CDM baseline. All results are derived from a joint analysis using DESI DR1, Planck 2018, and BBN constraints. The errors are presented at $68\%$ confidence except for the upper bounds that are at $95\%$ confidence level. The lower part of the table lists model-specific parameters, the minimum $\chi^2$ values, the statistical significance of the Hubble tension relative to SH0ES, and the Akaike Information Criterion (AIC) \cite{akaike2003new,burnham2004multimodel} differences with respect to $\Lambda$CDM.}
    \label{table:kessence_cosmopars}
\end{table*}

An important feature of the K-essence scenarios is the systematic shift in the cold dark matter density $\omega_{\rm dm}$ compared to the $\Lambda$CDM baseline. 
As shown in Table~\ref{table:kessence_cosmopars}, all K-essence models prefer values of $\omega_{\rm dm}$ slightly below the $\Lambda$CDM determination from \textit{Planck} 2018~\cite{Planck:2018vyg}, with best-fit shifts of order $\Delta \omega_{\rm dm} \sim -0.0015$ (about 1--1.5$\sigma$). 
This reduction compensates the additional early-time scalar contribution that mimics a radiation component, ensuring consistency with CMB acoustic peaks and BAO distances. 
The effect is robust across the kinetic, quadratic, and exponential cases, indicating that it is primarily driven by the modified background dynamics rather than model-specific potential effects.

\begin{figure}[H]
\centering
\includegraphics[width=0.6\linewidth]{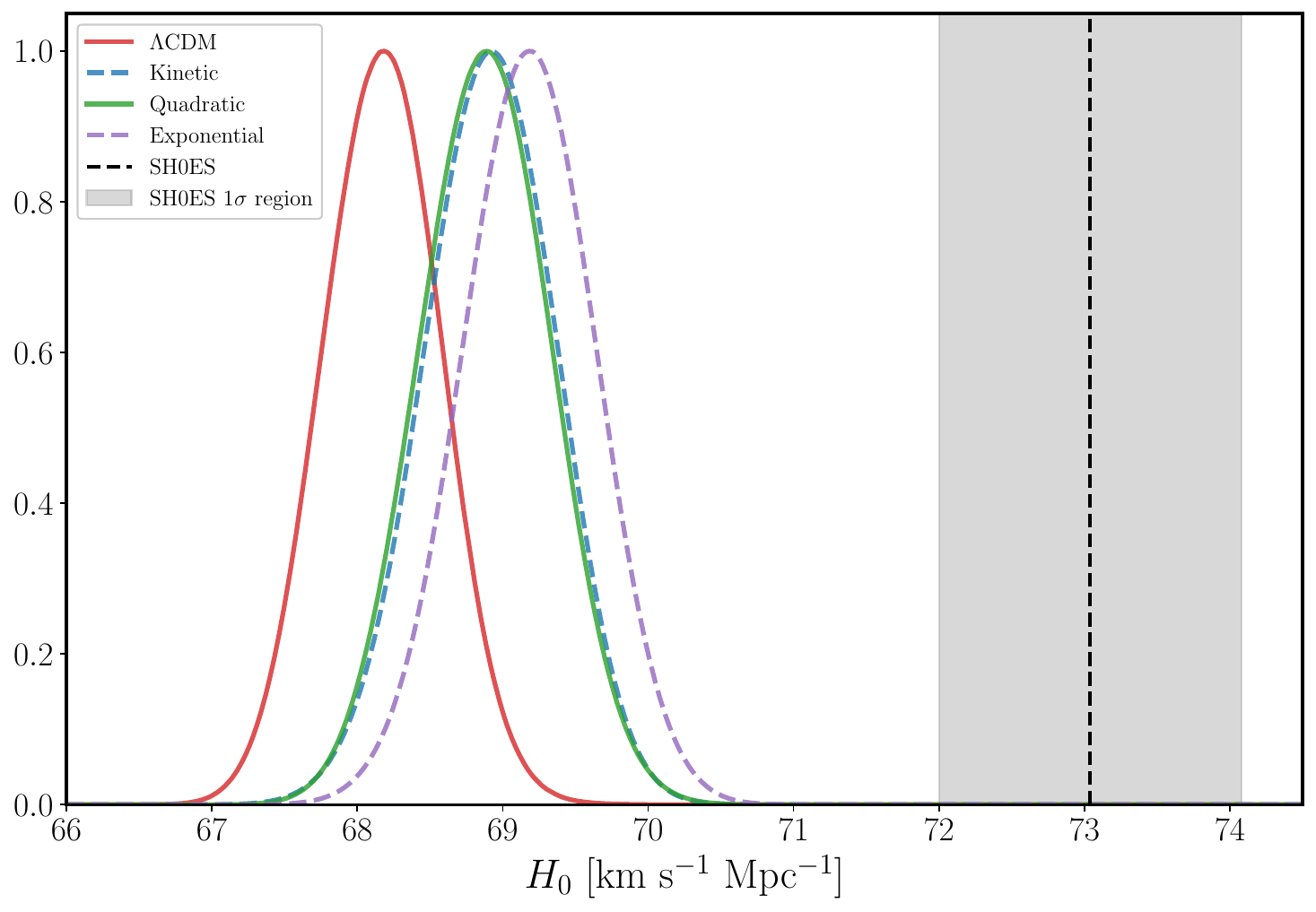}
\caption{Posterior probability function of $H_0$ 
for the three K-essence models and $\Lambda$CDM given the
BBN+Planck+DESI observations. 
The shaded region corresponds to the SH0ES measurement \cite{Riess:2021jrx} at $1\sigma$ confidence level. 
The field models shift the posterior towards higher $H_0$ reducing the tension to $3.66\sigma$ (kinetic), $3.68\sigma$ (quadratic), and $3.41\sigma$ (exponential), 
compared to $4.37\sigma$ in $\Lambda$CDM.}
\label{fig:allmodelsS8}
\end{figure}

To assess the impact of the K-essence scenarios on the Hubble tension, 
Figure~\ref{fig:allmodelsS8} shows the one-dimensional posterior probability distributions of $H_0$ 
for the three K-essence realizations and $\Lambda$CDM, using the combined 
BBN+Planck+DESI dataset.  
Relative to the $\Lambda$CDM baseline~\cite{Planck:2018vyg}, the kinetic and exponential models shift 
the posterior towards higher values, closer to the SH0ES measurement~\cite{Riess:2021jrx}, while the quadratic case essentially overlaps with the kinetic result due to its phenomenological degeneracy.  
Quantitatively, the Hubble tension is reduced from $4.37\sigma$ in $\Lambda$CDM to 
$3.66\sigma$ (kinetic), $3.68\sigma$ (quadratic), and $3.41\sigma$ (exponential).  
This shift reflects the generic effect of the scalar field’s early radiation-like phase on the expansion history, which propagates into higher inferred $H_0$ values.

Overall, the parameter constraints confirm that generalized K-essence models can fit current cosmological data at the same statistical level as $\Lambda$CDM, while introducing a mild but coherent shift in $H_0$ bringing its value closer to the region favored by late-Universe probes, while the systematic reduction in $\omega_{\rm dm}$ relative to \textit{Planck} 2018~\cite{Planck:2018vyg} emerges as a distinctive prediction of these scenarios.
This combination offers a possible observational handle to test K-essence against the standard $\Lambda$CDM paradigm in upcoming surveys.

The Akaike Information Criterion (AIC) given by 
\begin{equation}
    AIC = \chi^2_{min}+2k\,,
    \label{aic}
\end{equation}
where k is the number of parameters, is a criterion that evaluates the fit of the models to the data penalizing models with too many free parameters,
the smaller the value of the AIC the better.
From table \ref{table:kessence_cosmopars}, the field models fit better to the data even after penalizing them for the extra parameters, $(\theta)$ for the kinetic model, $(\theta, m_\phi)$ for de quadratic potential and $(\theta, V_0, \lambda)$ for the exponential. The purely kinetic model is the better suited according to the criterion.

\begin{figure*}[ht]
\centering
\begin{subfigure}[b]{0.32\textwidth}
\includegraphics[width=\textwidth]{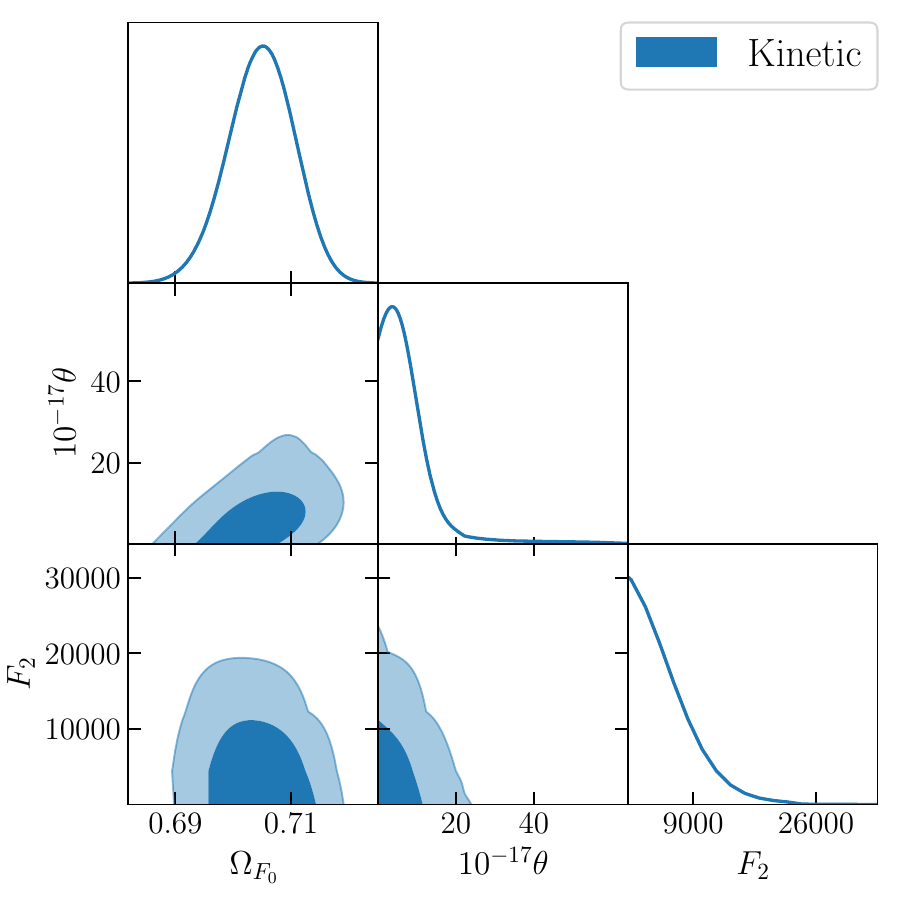}
\caption{Purely kinetic model.}
\label{fig:cinetico_params}
\end{subfigure}
\begin{subfigure}[b]{0.32\textwidth}
\includegraphics[width=\textwidth]{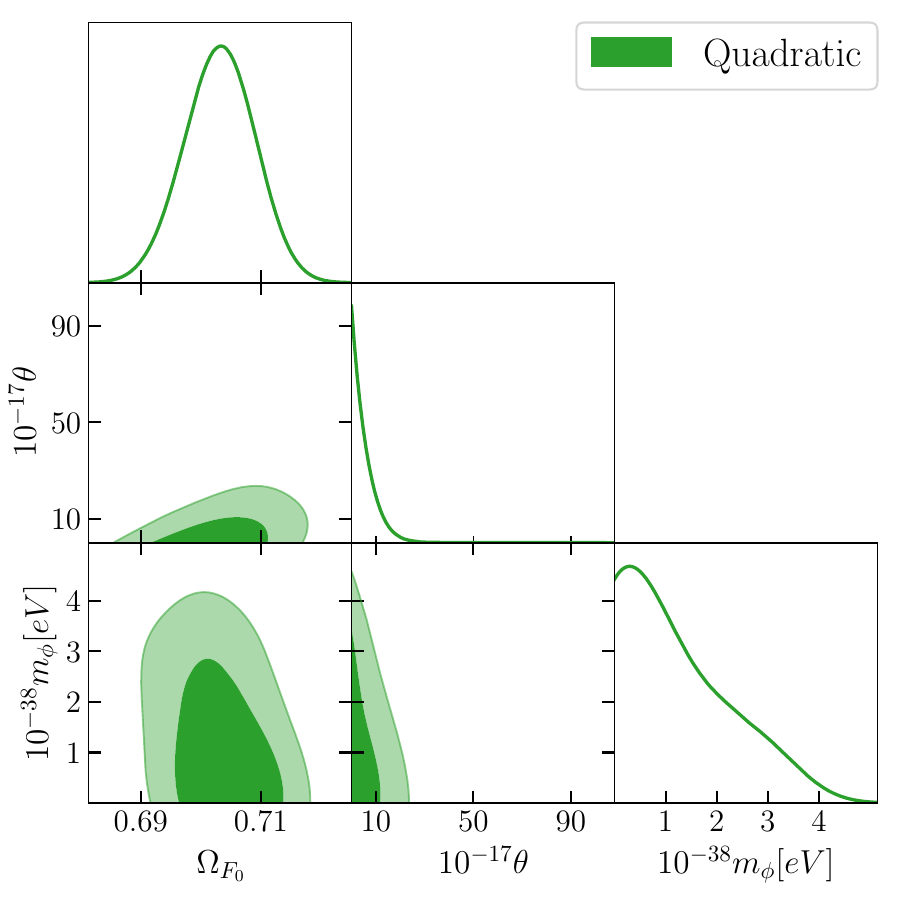}
\caption{Quadratic potential.}
\label{fig:cuadratico_params}
\end{subfigure}
\begin{subfigure}[b]{0.32\textwidth}
\includegraphics[width=\textwidth]{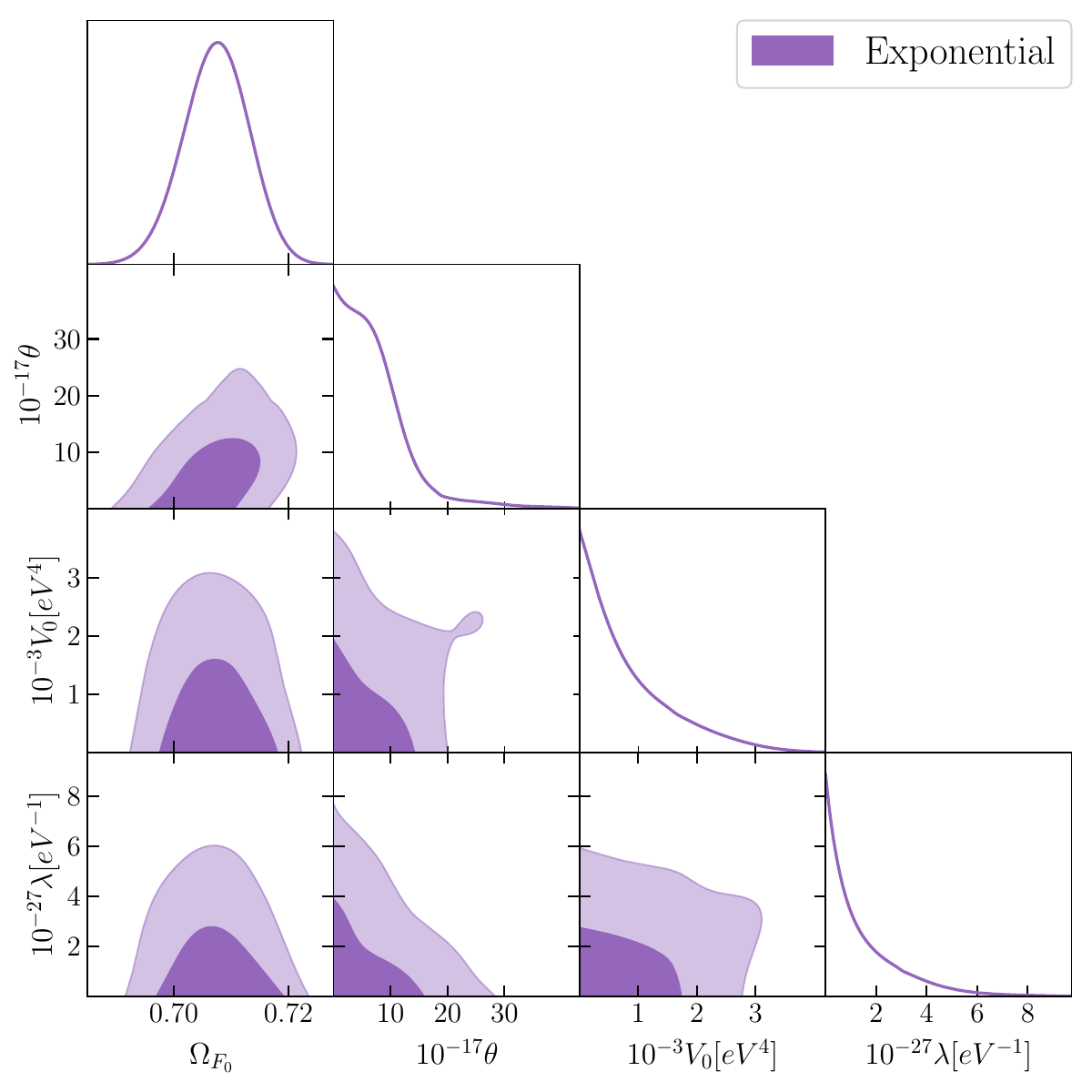}
\caption{Exponential potential.}
\label{fig:exponencial_params}
\end{subfigure}
\caption{Two-dimensional marginalized posterior distributions for the model-specific parameters in the three K-essence scenarios, compared to $\Lambda$CDM. Contours correspond to 68\% and 95\% confidence levels.}
\label{fig:posteriores_modelos_params}
\end{figure*}

Figure~\ref{fig:posteriores_modelos_params} displays the marginalized posterior distributions for the model-specific parameters in each K-essence scenario:  
\((\theta, \Omega_{F_0}, F_2)\) for the purely kinetic case,  
\((\theta, \Omega_{F_0}, m_\phi)\) for the quadratic potential,  
and \((\theta, \Omega_{F_0}, V_0, \lambda)\) for the exponential potential.  
A key difference between the purely kinetic model and its extensions 
is that in the purely kinetic, the coefficient \(F_2\) can be related explicitly to the matter density, since in this case the relations \eqref{relationF2_F0}
hold, allowing a direct mapping between the scalar Lagrangian and the cosmological density parameters.  
For the quadratic model, the posterior confirms that viable masses are restricted to \(m_\phi \lesssim 10^{-38}\,\mathrm{eV}\), consistent with the energetic bound in Section~\ref{sec:data}, which renders the potential dynamically irrelevant.  
In the exponential case, the posteriors admit a wide range of \((V_0,\lambda)\), with only a mild preference for non-zero values, and thus do not yield sharp constraints under current data.

The physical interpretation of our results highlights the central role of the kinetic sector in generalized K-essence cosmologies.  
In all scenarios considered, the constant term \(F_0\) provides the dominant contribution to the dark energy density today, effectively acting as a cosmological constant.  
This implies that the late-time acceleration of the Universe is predominantly driven by the non-canonical kinetic term, while scalar potentials play only a secondary role.  
The scalar potentials explored here mainly modulate the perturbation evolution, without altering the overall mechanism responsible for cosmic acceleration.

The quadratic potential case illustrates this point clearly.  
The requirement of an ultralight mass, \(m_\phi \le 10^{-38}\,\mathrm{eV}\), ensures numerical stability as well as a reasonable field density at late times,
but simultaneously renders the potential dynamically irrelevant throughout the cosmic history.  
As shown in Figure~\ref{fig:posteriores_modelos_x01}, the quadratic model essentially mimics the purely kinetic case, producing nearly indistinguishable predictions for both background and perturbations.  
Small residual differences in the dynamics remain, but these are well below the sensitivity of current observations.

The exponential potential, on the other hand, introduces a larger parameter space through \((V_0, \lambda)\), which in principle allows for richer phenomenology.  
Nevertheless, our constraints indicate no clear preference for a non-vanishing potential.  
The posterior distributions in Figure~\ref{fig:posteriores_modelos_params} remain broad and compatible with the purely kinetic limit, underscoring the robustness of the kinetic scenario.  
In this sense, the additional degrees of freedom in the exponential case do not lead to significant improvements in fit quality or distinct observational signatures under present data.

Across all models, the most relevant effects arise from the early-time radiation-like phase of the scalar field.  
As shown in Figure~\ref{fig:cinetico_Neff}, this phase temporarily increases the effective number of relativistic species, modifies the cold dark matter abundance, and leaves imprints on the growth of structure.  
Consistently, the two-dimensional posteriors in Figure~\ref{fig:posteriores_modelos_x01} show that all scenarios predict a systematic reduction in \(\omega_{\rm dm}\) compared to \(\Lambda\)CDM, compensating the scalar contribution and preserving the CMB acoustic scale and BAO distances.  
At the same time, the models induce coherent shifts toward higher \(H_0\), as illustrated in Figures~\ref{fig:posteriores_modelos_x01} and \ref{fig:allmodelsS8}, helping to partially ease the tension between SH0ES and CMB measurements.  
These signatures are not tied to the details of the potential, but rather stem from the universal features of the non-canonical kinetic sector.
As k-essence models  tend to increase the $H_0$ tension \cite{Lee:2022cyh,Colgain:2025nzf}, we speculate that this reduction might come from the extra radiation at  the early Universe as seen in other works \cite{Ben-Dayan:2023rgt,Ben-Dayan:2023htq}

Taken together, these findings reinforce the conclusion that the kinetic core of K-essence is the essential driver of its cosmological phenomenology.  
Potentials can be added without spoiling consistency with data, but they do not significantly alter the predictions within current observational precision.  
The distinguishing features of K-essence therefore lie in its early-time behavior and its impact on structure formation, providing a simple yet testable framework for unifying dark matter and dark energy.

\section{Conclusions}
\label{sec:conclusions}

In this work, we have analyzed a family of generalized K-essence models designed to unify dark matter and dark energy within a single scalar field framework.  
Building on the purely kinetic quadratic model \cite{Scherrer:2004au}, we studied two extensions with quadratic and exponential scalar potentials respectively. We evaluated whether such additions improve the phenomenological viability or yield distinctive observational signatures.  
The Scherrer's Lagrangian (\ref{scherrer}) is highly general, since any non-canonical Lagrangian $F(X)$ that possesses a minimum can be approximated by its expansion around that point.  
This universality has been emphasized in previous studies \cite{Chimento:2003ta,Bose:2008ew,De-Santiago:2011aka}.
Our results further explore this framework by confronting these scenarios with current cosmological datasets.

A further assessment of model performance was carried out using the Akaike Information Criterion (AIC), with $\Lambda$CDM taken as the reference.  
All K-essence scenarios yield lower AIC values.
Given the definition of the AIC parameter (Eq. \eqref{aic}),
this implies that the K-essence models achieve smaller $\chi^2$ values than
$\Lambda$CDM. Moreover, the reduction in $\chi^2$ exceeds
the penalty associated with the additional model parameters. Therefore, despite the increased model complexity, the net decrease in AIC indicates a modest statistical preference for the K-essence scenario. 
Among them, the purely kinetic Scherrer solution provides the best trade-off between fit quality and complexity, with $\Delta$AIC $\simeq -3$ compared to $\Lambda$CDM, while the addition of potentials seems to be counterproductive in terms of the statistical preference of the models.

These results, summarized in Table~\ref{table:kessence_cosmopars}, confirm that generalized K-essence scenarios
are able to  reproduce the cosmological observations (CMB, BAO and BBN) with roughly the same level of precision as $\Lambda$CDM, while unifying the dark sector under a single scalar degree of freedom.

We have shown that all models reproduce the expected sequence of cosmological epochs: an initial radiation-like phase, a matter-dominated era, and late-time accelerated expansion.  
In the purely kinetic case, this behavior arises naturally from the non-canonical structure and the constant offset $F_0$, which plays the role of a cosmological constant.  
The exponential potential preserves this structure while producing mild modifications at late times, whereas the quadratic potential requires an ultralight scalar mass of order $10^{-38}$~eV, rendering it dynamically indistinguishable from the kinetic scenario.

Future investigation should be made regarding the possibility that the small mass scale found in the quadratic model might be at odds with current observations. In the canonical field dark matter under a quadratic potential, the observational bounds  require $m_\phi \gtrsim 10^{-22}\rm{eV}$. These arise from suppression in the structure formation and the possibility to form galactic halos. In this model, the dark matter behavior arises primary from the kinetic sector and  the role of the small mass is not yet clear. A direct study of the halo models with non-canonical scalar fields should be made.
At the same time, the matter power spectrum comparison in Fig.~3 shows that the models with scalar potentials are strongly constrained by large scale structure data.

A key outcome is the scalar field’s early-time relativistic behavior, which enhances the radiation density and contributes to the effective number of relativistic species.  
We explicitly computed this contribution and identified the parameter region $\theta \lesssim 19.7 \times 10^{-17}$ as required to satisfy BBN and \textit{Planck} constraints.  
This bound applies uniformly across all models, providing a robust early-Universe constraint on the dynamics of generalized K-essence.  

From the parameter inference perspective, all scenarios remain consistent with \textit{Planck} 2018, DESI DR1 BAO, and BBN data.  
They systematically predict slightly lower $S_8$ and higher $H_0$ relative to $\Lambda$CDM, leading to modest shifts toward the region favored by SH0ES and weak lensing surveys.  
The kinetic and exponential cases reduce the Hubble tension to the $\sim 3\sigma$ level, while the quadratic potential remains redundant with the kinetic limit.  
In addition, all models prefer values of the cold dark matter density $\omega_{\rm dm}$ below the $\Lambda$CDM baseline, compensating for the scalar’s relativistic contribution at early times and preserving consistency with CMB acoustic peaks and BAO distances.  

Although current data do not provide a statistical preference for K-essence over $\Lambda$CDM, the coherent parameter shifts and theoretical robustness of the framework highlight its potential as a unifying description of dark matter and dark energy.  
Recent work has suggested that extensions involving direct couplings to matter may further alleviate the $H_0$ and $S_8$ tensions by modifying the effective gravitational coupling at different epochs~\cite{HosseiniMansoori:2024pdq}.  
Future observations with next-generation galaxy surveys such as DESI DR2~\cite{tr6y-kpc6}, Euclid~\cite{ refId0}, and the Vera C. Rubin Observatory~\cite{Ivezić_2019}, together with upcoming CMB experiments such as CMB-S4~\cite{Abazajian:2019eic} and the Simons Observatory~\cite{Ade_2019}, will be decisive to test these models beyond the linear regime and to search for distinctive signatures in gravitational lensing, clustering, and non-linear structure formation.

\funding{EM was supported by SECIHTI grant 815225 and by SECIHTI project CBF2023-2024-589.}
 
 \bibliographystyle{JHEP}  
 \bibliography{refs.bib}

\end{document}